\documentclass[aps,showpacs,preprint,epsf,epsfig,showkeys, nofootinbib,longbibliography]{revtex4-2}
\usepackage{mathrsfs}
\usepackage{dcolumn}
\usepackage{bm}
\usepackage{hyperref}
\usepackage{epsfig,graphicx,color,appendix}
\usepackage{placeins}
\usepackage{multirow}
\usepackage{tabularx}
\usepackage{capt-of}
\usepackage{caption}
\usepackage[countmax]{subfloat}

\usepackage{threeparttable}
\usepackage{amssymb}
\usepackage{array}
\usepackage{amsmath}
\usepackage{rotating,tabularx}
\usepackage{float}
\usepackage[english]{babel}

\begin{document}
\title{Quark-diquark effective mass formalism for heavy baryon spectroscopy}
\author{\textbf{Binesh Mohan}}
\email{bineshmohan96@gmail.com}
\author{\textbf{Rohit Dhir}}
\email[Corresponding author: ]{dhir.rohit@gmail.com}
\affiliation{Department of Physics and Nanotechnology,\\SRM Institute of Science and Technology, Kattankulathur-603203, Tamil Nadu, India.}

	\long\def\symbolfootnote[#1]#2{\begingroup%
		\def\thefootnote{\fnsymbol{footnote}}\footnote[#1]{#2}\endgroup}
	\def\lsim{ {\ \lower-1.2pt\vbox{\hbox{\rlap{$<$}\lower5pt\vbox{\hbox{$\sim$}
			}}}\ } }
	\def\gsim{ {\ \lower-1.2pt\vbox{\hbox{\rlap{$>$}\lower5pt\vbox{\hbox{$\sim$}
			}}}\ } }
	
	\vskip 1.5 cm
	\small
	\vskip 1.0 cm

\date{\today}
\begin{abstract}
We develop a quark–diquark effective mass formalism  for heavy-flavor baryon spectroscopy and apply it to the $J^P = \tfrac{1}{2}^+$ and $J^P = \tfrac{3}{2}^+$ 
spectra across the singly, doubly, and triply heavy sectors. The analysis is carried out in two complementary scenarios: Scenario~I treats all quark–quark diquark 
channels dynamically, while Scenario~II restricts the dynamics to scalar and axial-vector diquarks, providing a more selective and physically transparent description. Constituent quark masses, effective diquark masses, and chromomagnetic couplings are extracted from known heavy-baryon masses, with the couplings determined solely by the quark content of each state and no sector-dependent adjustment introduced. A mass-dependent binding term is implemented to account for spin-independent chromoelectric effects and to describe the transition from chromomagnetic to color-Coulomb dominance across the light-to-heavy quark regime, ensuring consistency with heavy-quark spin symmetry in the heavy-quark limit. The resulting predictions are in good agreement with available experimental measurements and lattice QCD results across both charm and bottom sectors. The extracted diquark parameters remain stable across all heavy-flavor sectors, establishing the present framework as a symmetry-constrained spectroscopic baseline for heavy-baryon structure.

\end{abstract}
	
\maketitle
\newpage

\section{Introduction}
\label{Sec1}

Recent years have seen remarkable progress in the spectroscopy of heavy flavor baryons, largely driven by LHCb and CMS. Nearly the full ground-state singly heavy baryon spectrum has now been measured~\cite{ParticleDataGroup:2024cfk}. A major milestone was the observation of $\Xi_{cc}^{++}$ and its isospin partner $\Xi_{cc}^{+}$~\cite{LHCb:2017iph, LHCb:2019epo, LHCb:2026pxn, LHCb:2021eaf, LHCb:2019gqy}. Searches for the doubly heavy baryons $\Xi_{cb}^{+(0)}$, $\Omega_{cb}^{0}$, and $\Omega_{cc}^{+}$ are ongoing~\cite{LHCb:2022fbu, LHCb:2021xba, LHCb:2021rkb, LHCb:2020iko}. The LHCb Upgrade II, with higher luminosity, is expected to reveal additional states in the HL-LHC era~\cite{Vagnoni:2025qfv}, while future $e^+e^-$ colliders promise detailed studies of doubly and triply heavy baryons~\cite{Zhan:2023jfm, FCC:2018evy, CEPCStudyGroup:2018rmc, CEPCStudyGroup:2018ghi, ILC:2007bjz}. The discovery of these heavy baryons provides crucial insight into heavy-quark dynamics and quantum chromodynamics (QCD) confinement, bridging conventional and exotic multiquark systems.

A pressing theoretical question now confronts this experimental richness: can the same quark correlations that govern conventional baryon structure also organize the spectroscopy of tetraquark and pentaquark states? The observation of the open-flavor doubly heavy tetraquark $T_{cc}(3875)^+$ with a precisely measured mass of $3874.74(10)$~MeV~\cite{LHCb:2021vvq}, together with a growing set of tetraquark candidates~\cite{LHCb:2024smc, LHCb:2022sfr, Jiang:2024lsr}, has strengthened the empirical link between conventional hadron spectroscopy and multiquark dynamics. More broadly, a wide range of manifestly exotic states has now been reported. These include charged hidden-flavor tetraquarks, such as $Z_c(3900)$ and their bottom counterparts, as well as fully heavy resonant structures, including the di-$J/\psi$ resonances $X(6600), X(6900)$, and $X(7100)$ observed by CMS, LHCb, and ATLAS~\cite{CMS:2023owd, LHCb:2020bwg, ATLAS:2023bft}. The emerging hidden-charm pentaquark spectrum~\cite{Belle:2025pey, LHCb:2022ogu, LHCb:2021chn, LHCb:2020jpq, LHCb:2019kea, LHCb:2015yax} further reinforces the view that multiquark hadrons are a persistent feature of the QCD spectrum. These observations have motivated numerous theoretical approaches, including potential models, relativistic quark models, QCD sum rules, chiral quark models, and lattice QCD~\cite{Ferretti:2019zyh, Yin:2019bxe, Farhadi:2023ucs, MoosaviNejad:2020nsl, Kim:2021ywp, Santopinto:2014opa, Faustov:2021qqf, Ebert:2011kk, Ebert:2002ig, Jiang:2014ena, Yao:2018ifh, Shah:2016mig, Shah:2016vmd, Shah:2017jkr, Shah:2017liu, Zhang:2021yul, ShekariTousi:2024mso, Aliev:2014lxa, Aliev:2012tt, Aliev:2012iv, Liu:2007fg, Ortiz-Pacheco:2023kjn, Roberts:2007ni, Bahtiyar:2022nqw, Bahtiyar:2020uuj, Mohanta:2019mxo, Mathur:2018rwu, Mathur:2018epb}. While these approaches have yielded valuable sector-specific insights, none provides a single framework in which diquark properties are extracted from and validated against the current baryon spectrum and subsequently applied to multiquark systems. Establishing such a calibration pipeline motivates the present work.

Within this context, diquark correlations~\cite{Barabanov:2020jvn, Jaffe:2004ph, Anselmino:1992vg} provide a robust, quantitatively testable description of baryon structure and serve as a natural bridge between conventional and exotic hadrons~\cite{Liu:2019zoy, Karliner:2017qhf, Esposito:2016noz, Lebed:2016hpi, Chen:2016qju}. Recent CMS limits on quark substructure down to $10^{-20}$~m~\cite{CMS:2026ecv} preclude spatial smearing at TeV scales, justifying the contact-interaction limit in our chromomagnetic interaction and reinforcing the treatment of the diquark as a dynamically compact cluster. This interpretation is further supported by flavor-decomposition analyses of nucleon electromagnetic form factors at high momentum transfers~\cite{Barabanov:2020jvn, Cates:2011pz}. Additional evidence comes from heavy-baryon production mechanisms. For instance, double-charm production at LHCb proceeds predominantly through the formation of a compact heavy diquark that subsequently hadronizes~\cite{Ma:2025ito, Trunin:2016uks, Chen:2014frw, Likhoded:2009zz}.

Recent analysis using the Gaussian expansion method have identified clear diquark correlations within baryons. In particular, the scalar-isoscalar diquark is found to have lower energy and a more compact spatial extent, consistent with the notion of a ``good diquark''~\cite{Zhu:2025qfi}. For $QQ$-$q$ systems, increasing the heavy-quark mass reduces the diquark separation and enhances the diquark clustering effect, supporting a compact heavy diquark picture. The quark-diquark framework has become standard for describing triply heavy baryons~\cite{Faustov:2021qqf}, constructing unified Regge trajectories for heavy-heavy systems~\cite{Xie:2024lfo, Chen:2023djq}, and formulating chiral effective theories of diquarks~\cite{Kim:2024tbf}. Collectively, these developments show that the diquark configuration, by capturing dominant color-spin correlations and respecting HQSS, is a natural and necessary building block for extending QCD dynamics from conventional baryons to multiquark exotics.

Thus, calibrating diquark properties directly from the baryon spectrum is a phenomenological necessity, not a mere modeling simplification. Motivated by this, we develop a quark-diquark effective mass formalism (QDEMF) in two complementary scenarios. Scenario I treats all possible diquark combinations arising from quark-quark interactions, providing a dynamically inclusive description. Scenario II instead fixes the diquark spin-flavor structure, restricting the dynamics to flavor-antisymmetric scalar and flavor-symmetric axial-vector diquarks, thereby furnishing a more selective and physically transparent interaction picture. To account for the changing dynamical regime between light and heavy quarks, we introduce a mass-dependent binding energy term~\cite{Karliner:2014gca}. This term captures short-range chromoelectric effects and the gradual transition from chromomagnetic dominance to spin-independent color-Coulomb dynamics in accordance with heavy-quark spin symmetry (HQSS).

The present framework yields several concrete outcomes. First, it provides predictions for the low-lying heavy-baryon spectrum across singly, doubly, and triply heavy sectors within a single parameter-light formulation. Second, it enables a systematic extraction and analysis of effective diquark masses and flavor content within the two complementary scenarios introduced above, with scalar and axial-vector contributions explicitly disentangled across all heavy-flavor sectors. Third, the framework provides a unified description of both scalar and axial-vector diquarks together with a transparent accounting of hyperfine (spin-spin) splittings. The characteristic $1/(m_i m_j)$ chromomagnetic scaling arises naturally from the underlying one-gluon exchange (OGE) interaction and is reproduced consistently across all flavor sectors as a direct consequence of the dynamics. Consequently, the correct HQSS degeneracy structure emerges automatically in the heavy-quark limit, rather than being imposed as an external constraint.

The chromomagnetic couplings are determined entirely by the quark content of each state, with no sector-dependent adjustment introduced at any stage, so that the extracted diquark masses and hyperfine couplings act as stable predictive parameters across the full heavy-baryon spectrum. Within the present work we restrict attention to baryon spectroscopy, using the resulting symmetry patterns as the primary internal validation of the quark-diquark framework.\footnote{The broader applicability of the calibrated diquark parameters to multiquark spectroscopy is explored in a companion study~\cite{Mohan:2026blk}, where the extracted mass scales and hyperfine couplings are employed without refitting to construct tetraquark and pentaquark spectra across multiple flavor sectors.}

The paper is organized as follows. Sec.~\ref{Sec2} presents the QDEMF, including the binding energy prescription. Numerical results and discussions are given in Sec.~\ref{Sec3}. Sec.~\ref{Sec4} summarizes the conclusions. Appendices~\ref{A1}-\ref{A3} contain the diquark and baryon wave functions, and mass relations.
\section{Methodology}
\label{Sec2}
\subsection {Quark-diquark effective mass formalism}
\label{qDq}

The quark-diquark model proposes that the valence quarks within a baryon may favor a spatially proximate arrangement rather than random motion. This simplification allows for the assumption that a baryon is a two-body bound state of a diquark and a quark, effectively simplifying the dynamics of the three-body system \cite{Anselmino:1992vg}. Similar to quarks, a diquark, the bound state of two quarks, is a colored object confined within a baryon. Since quarks transform like a color $SU(3)$ triplet, a diquark's color structure can be either an antitriplet ($\overline{3}_c$) or a sextet ($6_c$), resulting from the direct product:
\[3_c \otimes 3_c = \overline{3}_c \oplus 6_c.\]
The antitriplet color state, with a coefficient of $-2/3$, indicates an attractive interaction where two quarks are bound to form a stable configuration. Conversely, gluon exchange is repulsive in the sextet state due to its $+1/3$ coefficient \cite{Ali:2017jda, Ali_Maiani_Polosa_2019, Karliner:2014gca, Jaffe:2004ph, Cahill:1987qr}. Consequently, diquarks must be in a $\overline{3}_c$ state to form a color singlet baryon when combined with a $3_c$ quark. This allows for treating a baryon as a two-body system where a quark within the baryon perceives the diquark as an antiquark, behaving like a $\overline{3}_c$ state, analogous to the meson system \cite{Mutuk:2021zes}.

Remarkably, a diquark in a low-lying ($l=0$) configuration can exhibit a total spin of either 1 or 0, similar to ground-state mesons. However, unlike mesons, ground-state diquarks possess positive parity and are thus categorized as axial-vector ($A$) and scalar ($S$) diquarks with $J^P = 1^+$ and $0^+$, respectively. In an axial-vector diquark, the quark spins are aligned parallel, whereas in a scalar diquark, they are antiparallel, leading to a stronger binding interaction in the latter \cite{Ali:2017jda, Jaffe:2004ph}.

Upon including $SU(3)$ flavor degrees of freedom, the $SU(6)$ spin-flavor representation of the diquark state arises from the direct product as follows:\footnote{Here, $f$ and $s$ refer to flavor and spin degrees of freedom, respectively.}
\begin{equation*}
\begin{split}
    6 \otimes 6 & = \overline{15} \oplus 21 \\
    & = (\overline{3}_f \otimes 3_s) \oplus (6_f \otimes 1_s) \oplus (6_f \otimes 3_s) \oplus (\overline{3}_f \otimes 1_s).
\end{split}
\end{equation*}
However, according to the Pauli principle, a diquark prefers a symmetric 21-dimensional representation of $SU(6)$ over the antisymmetric 15-dimensional representation to form a totally antisymmetric baryon wave function \cite{Barabanov:2020jvn}. A diquark composed of two identical quark flavors can only exist in a spin-1 state, while a diquark composed of two different quark flavors can appear in both spin-1 and spin-0 states. The explicit $SU(6)$ wave functions for these diquark configurations are provided in Appendix \ref{A1}.

Consequently, baryon wave functions can be expressed in terms of the wave functions of individual quarks and diquarks. Decuplet ($J^P=\frac{3}{2}^+$) baryons are composed solely of axial-vector diquarks, while octet ($J^P=\frac{1}{2}^{(\prime)+}$) baryons can contain both axial-vector and scalar diquarks, which are equally weighted in the limit of $SU(6)$ symmetry. The resulting quark-diquark wave functions of $SU(3)$ baryons are given in Appendix \ref{A2}. These wave functions are normalized to reproduce the three-quark wave functions of baryons \cite{Lichtenberg:1975ap}. Therefore, the probability of a diquark in a baryon being in an axial-vector or scalar state is dictated by the Clebsch-Gordon (CG) coefficients involved in the limit of exact $SU(6)$ symmetry \cite{Khanna:1983qwl, Anisovich:2010wx}. Thus, using the baryon wave function expressed as a quark-diquark configuration, one can obtain the baryon mass expressions in terms of the masses of quarks and diquarks. Both axial-vector and scalar diquarks contribute to the total mass of the baryon states, with their overall contributions determined by the corresponding CG coefficients associated with the wave functions. Appendix \ref{A3} lists the baryon mass expressions obtained from quark-diquark wave functions.

The baryon mass, though fundamentally governed by the QCD Hamiltonian, is often estimated using a non-relativistic three-quark model. In this approach, a unitary transformation projects the QCD Hamiltonian onto a parametrized mass operator acting on quark-level spin-flavor wave functions \cite{Morpurgo:1999mr, Dillon:2009pf}. This operator effectively encodes key QCD dynamics, including confinement, chiral symmetry breaking, gluon exchange, and diquark correlations. Within this framework, the spin-flavor structure dictated by $SU(6)$ symmetry determines the relative contributions of scalar and axial-vector diquarks, weighted by CG coefficients. These coefficients influence the operator’s expectation values, thereby affecting the baryon mass and hyperfine splittings \cite{Khanna:1983qwl}. Importantly, all three diquark pairings contribute via quark-quark interactions mediated by OGE.

We introduce QDEMF as a refined framework for estimating baryon masses by incorporating diquark correlations directly into the effective mass scheme (EMS)~\cite{Mohan:2022sxm, Hazra:2021lpa, Dhir:2013nka, DeRujula:1975qlm}. In the standard EMS, all spin-independent effects of confinement and color-Coulomb dynamics are absorbed into renormalized constituent quark masses, while strong hyperfine interactions are treated explicitly via spin–spin correlations. In the present QDEMF, we retain this hyperfine structure but depart from the standard scheme by computing the spin-independent binding energy contribution explicitly and separately, rather than absorbing it into the constituent masses \cite{Morpurgo:1989my, Morpurgo:1999mr}. This separation allows the color-Coulomb dynamics to be tracked across the light-to-heavy quark transition without conflating spin-independent and spin-dependent contributions at the level of the effective diquark mass. This approach models baryons as three constituent quarks interacting via spin-dependent OGE,
\begin{equation}
\label{eq1} 
M_B = \sum_{i=1}^{3}{m_i} + \sum_{i<j}b_{ij}\bm{s_i.s_j},
\end{equation}
where $m_i$ are constituent masses of the $i$th quark in a baryon. The spin operators of the $i$th and $j$th quarks are designated by $\bm{s_i}$ and $\bm{s_j}$ correspondingly. The strong hyperfine interaction term, $b_{ij}$, is given by
\begin{equation}
\label{eq2}
 b_{ij} =\frac{16\pi\alpha_s}{9m_im_j}\langle\psi|\delta^3(\vec{r})|\psi\rangle,
\end{equation}
where $\psi$ and $\alpha_s$ represent the baryon wave function at the origin and the strong coupling constant, respectively. Unlike EMS, QDEMF recasts the baryon as a bound state of a diquark and a quark. This two-body perspective captures dominant quark-quark interactions and offers a more physically motivated though simplistic description, particularly for systems where diquark clustering is favored. In QDEMF, the baryon mass is expressed as,
\begin{equation}
\label{eq_diquark_pairs}
M_B = \frac{1}{3}\sum_{i<j} \left( m_{ij}^\delta + m_k \right) + \sum_{i<j} b_{ij}\, \bm{s_i.s_j},
\end{equation}
where $m_{ij}^\delta = m_i + m_j$ denotes the diquark formed by quarks $i$ and $j$. The spin-spin expectation values for the two-quark subsystem are,
\begin{align}
\label{exp_spin_v}
\langle \bm{s_i.s_j} \rangle_{S=1} &= +\tfrac{1}{4}, \\
\label{exp_spin_s}
\langle \bm{s_i.s_j} \rangle_{S=0} &= -\tfrac{3}{4}.
\end{align}

The implementation of constituent diquark and quark masses within the QDEMF mirrors the leading-order hierarchy found in chiral perturbation theory and symmetry-based mass expansions \cite{Morpurgo:1989my, Dillon:1995qw}. Under this organizing principle, the mass spectrum is primarily dictated by the internal scales of the diquark and the spectator quark, alongside the dominant color-spin interactions between them. The empirical success of the Gell-Mann--Okubo formula \cite{Durand:2001zz, Durand:2001sz, Dillon:2002ks} supports this approach, suggesting that complex many-body effects and higher-order nonlinear operators remain subdominant to the fundamental quark-diquark architecture. 

The strength of the hyperfine interaction is modulated by the color factor $\langle \vec{\lambda}_i \cdot \vec{\lambda}_j \rangle$, with $\vec{\lambda}_i$ representing Gell-Mann matrices acting on the color space of the $i$th quark. In mesons (quark-antiquark singlets), this factor is $-4/3$, while in baryons, for quark-quark pairs in a color antitriplet, it reduces to $-2/3$. This reduction reflects the weaker effective color-magnetic coupling within baryons, as first proposed by De Rújula, Georgi, and Glashow \cite{DeRujula:1975qlm}. By systematically incorporating diquark formation into the mass formula, QDEMF preserves key features of QCD-inspired dynamics while offering a compact and predictive tool for baryon spectroscopy. It is particularly well-suited for describing heavy baryons and exotic states where diquark structures are expected to dominate.

For the interaction between the diquark (in the color antitriplet $\overline{3}$) and the third quark (in the triplet $3$), the color factor for the hyperfine interaction is the same as in a meson. Since the quadratic Casimir for the $\overline{3}$ representation is $C_2(\overline{3}) = \frac{4}{3}$, the color factor becomes,\footnote{For $SU(N)$, the quadratic Casimir is $C_2(N) = \frac{N^2 - 1}{2N}$; hence, for $SU(3)$: $C_2(3) = C_2(\overline{3}) = \frac{4}{3}$.}
\begin{equation}\label{casimir}
\langle\vec{\lambda}_{\text{diquark}} \cdot \vec{\lambda}_k\rangle_{\text{singlet}} 
= \frac{1}{2}\left( C_2(\text{singlet}) - C_2(\overline{3}) - C_2(3) \right)
= \frac{1}{2} \left( 0 - \frac{4}{3} - \frac{4}{3} \right)
= -\frac{4}{3}.
\end{equation}
This matches the meson case, while in contrast, quark-quark color couplings in baryons yield a smaller factor of $-2/3$. Many traditional effective mass schemes absorb such color factors into the definition of $b_{ij}$. However, when reorganizing the baryon as a quark-diquark system, this color enhancement must be explicitly accounted for. 

In a three-quark baryon, hyperfine interactions arise from all three distinct quark-quark pairs contributing to the system's spin-spin coupling. When reformulated using the quark-diquark picture, only one pair is explicitly treated as a diquark, effectively capturing only one-third ($\frac{1}{3}$) of the total hyperfine contributions from the original three-body system. To compensate for this undercounting and ensure accurate reproduction of observed baryon hyperfine splittings, the hyperfine term in the effective diquark mass must be scaled by a factor of three. This scaling factor accounts for all three quark-quark pairs in the baryon's color-spin dynamics, integrating their collective effect into the diquark mass to maintain consistency with the complete three-quark description while providing a computationally tractable framework. Therefore, the effective diquark mass must be written as
\begin{equation}
\label{eq:diquark_mass_enhanced}
m_{ij}^{\mathscr{E}} = m_i + m_j + 3~b_{ij}\, \bm{s_i.s_j},
\end{equation}
where the factor of $3$ ensures consistency with the original three-body description. Therefore, the mass expressions for axial-vector ($1^+$) and scalar ($0^+$) diquarks follow from the spin-spin expectation values in Eqs. \eqref{exp_spin_v} and \eqref{exp_spin_s}:
\begin{align}
\label{axial}
M_{A(1^+)} &= m_i + m_j + 3\left(\frac{b_{ij}}{4}\right), \\
\label{scalar}
M_{S(0^+)} &= m_i + m_j - 3\left(\frac{3b_{ij}}{4}\right).
\end{align}

These refined mass expressions provide the input for the QDEMF, ensuring that both color and spin contributions are consistently treated within the baryon mass calculation. A key feature of this framework is its treatment of all quark-quark hyperfine interactions within the baryon via distinct diquark configurations, effectively absorbing the baryon's total hyperfine energy, approximated as equally distributed among the three quark pairs, into the diquark mass. This implies that strong hyperfine interactions exist solely between quarks within the diquark, rather than between the diquark and its residual quark. While this approach omits explicit quark-diquark hyperfine interactions, it substantially reduces parameters and computational complexity. This approximation proves well-suited for symmetric baryons and maintains reliability for heavy-light systems, where $b_{ij}$ values are typically small, as observed in EMS analyses \cite{Mohan:2022sxm}. Despite a potential slight overestimation of hyperfine effects in such instances, the impact remains minimal. This practical framework, mirroring similar methods in the literature that do not explicitly identify diquarks, offers a simplified yet effective means to describe exotic state dynamics. This specific representation of the quark-diquark model emphasizes the internal quark-quark interplay within the baryon, where diquark spin-flavor degrees of freedom remain unfixed. Neglecting explicit quark-diquark interactions, this framework is termed the ``quark-quark interaction picture" and is hereby denoted as ``Scenario I". 

In QDEMF, the experimental baryon masses reported by the PDG \cite{ParticleDataGroup:2024cfk} are used to extract constituent quark masses ($m_i$) and strong hyperfine interaction terms ($b_{ij}$) in a model-independent manner. These parameters are determined from known baryon mass relations, inherently capturing flavor-dependent effects, including isospin breaking. We adopt the isospin-broken quark masses and $b_{ij}$ values from our previous EMS-based analysis \cite{Mohan:2022sxm}, which focused on the light and charm sectors. This input ensures consistency with isospin-breaking effects in the computed diquark masses. Although the earlier work employed a fully symmetric three-quark framework, the $b_{ij}$ terms defined for mutual quark pairs retain their physical significance in the quark-diquark description as well. Since these hyperfine terms stem from two-body interactions, their values are largely insensitive to whether the framework is three-body (quark-quark-quark) or two-body (quark-diquark), provided the flavor configuration is preserved. In our case, the $b_{ij}$ values extracted from the strange and charm sectors based on experimental baryon masses were found to be nearly identical in both pictures. Hence, we continue to use the same EMS-based $b_{ij}$ inputs for consistency and simplicity.

For the bottom sector, the analysis presented here is entirely new. Due to the limited experimental data, particularly for baryons containing multiple bottom quarks, flavor symmetry (breaking) arguments are employed to estimate the relevant hyperfine terms. A detailed discussion of the specific input values for both charm and bottom sectors is provided in Sec. \ref{Sec3}.

\subsection*{Improved Scenario II}
\begin{figure}[]
    \centering
    \includegraphics[width=1.0\linewidth]{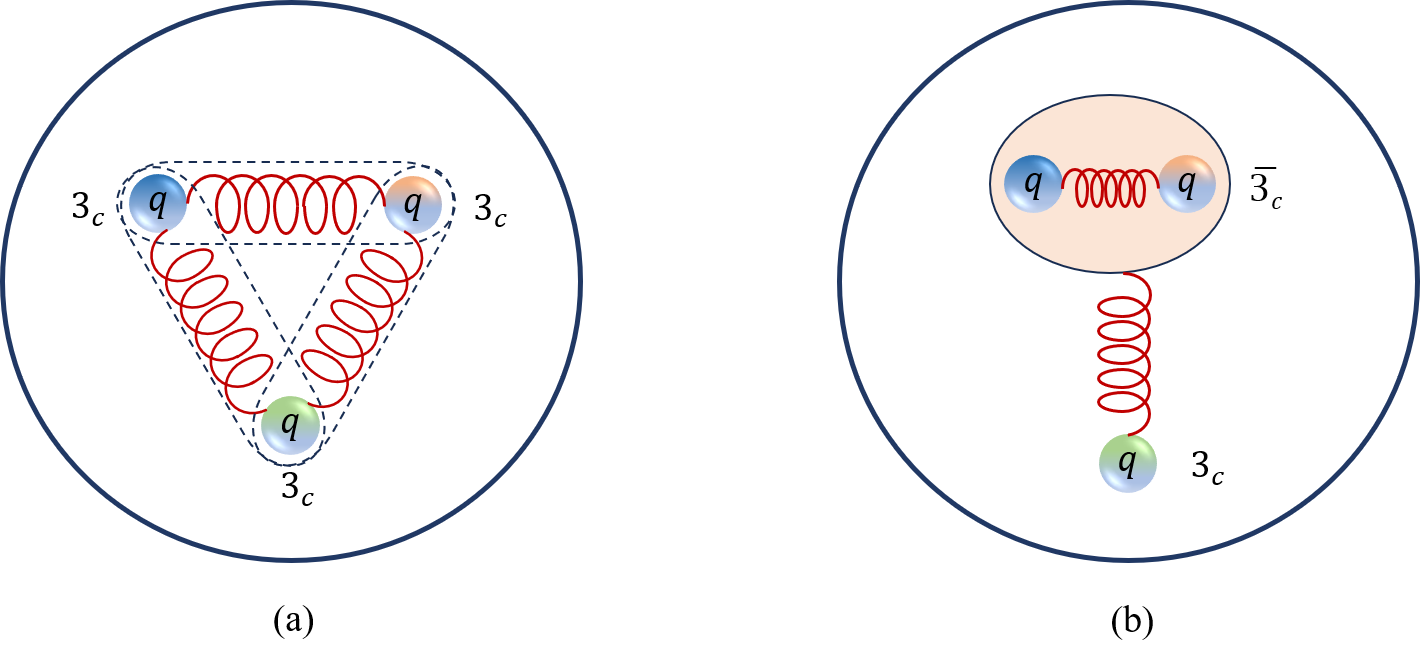}
    \caption{Two scenarios in QDEMF: (a) quark-quark interaction picture (Scenario I), (b)~quark-diquark interaction picture (Scenario II).}
    \label{Fig3}
\end{figure}

Alternatively, we introduce a refined quark-diquark model, termed the ``quark-diquark interaction picture'' and denoted as ``Scenario II'', which explicitly incorporates the quark-diquark interplay through the hyperfine interaction between the diquark and the third quark, alongside the intra-diquark quark-quark interaction. This approach enables a more complete treatment of spin-flavor correlations and effectively reduces the baryonic three-body system to a two-body framework, analogous to mesons, while preserving key dynamical features. The schematic representation of scenarios I and II in QDEMF is shown in Fig. \ref{Fig3}.
\FloatBarrier

Unlike Scenario I, where the total hyperfine energy is approximated as equally distributed among three quark pairs and absorbed into the diquark mass (using an effective factor of $3$), Scenario II systematically incorporates the color factor difference between quark-quark ($-2/3$) and quark-antiquark or quark-diquark ($-4/3$) couplings. This corrects for potential overestimations of hyperfine effects in heavy-light systems, especially where $b_{ij} \propto 1/(m_i m_j)$ is already suppressed and model uncertainties can be obscured by dominant quark mass contributions.

Although Scenario II introduces additional parameters, this formalism offers clear advantages: it decouples intrinsic diquark properties from emergent baryon dynamics, retains QCD-motivated structure, and allows scalable extensions to multiquark systems. In contrast, Scenario I, while practical for conventional baryons, becomes unwieldy when generalized to exotic configurations due to the proliferation of undetermined parameters and complex color-spin diagonalization. Scenario II addresses these limitations by isolating the relevant interactions, thus providing a more robust and analytically tractable framework for studying both conventional and exotic hadrons. Therefore, in QDEMF, the baryon mass in Scenario II is governed by the quark-diquark interaction originating from OGE, given  by
\begin{equation}
\label{Eq5} 
    M_B = m_{ij} + m_k +b_{\{ij\}k}~\bm{s_D.s_k},
\end{equation}
where $m_{ij}$ represents the effective mass of the diquark $D$ (with flavor content $ij$) and $m_k$ corresponds to the constituent mass of the third quark $k$, respectively. In this framework, the effective diquark mass, $m_{ij}$ can take one of two forms depending on the spin configuration of the quark pair. Specifically, $m_{\{ij\}}$ and $m_{[ij]}$ correspond to the axial-vector ($D^{\{ij\}}$) and scalar ($D^{[ij]}$) diquark masses in Scenario II, respectively. These diquark masses are obtained using their corresponding spin-spin expectation values given in Eqs. \eqref{exp_spin_v} and \eqref{exp_spin_s}. Therefore,
\begin{equation}  
\label{Eq10} 
		m_{\{ij\}} = m_i + m_j + \frac{b_{ij}}{4},
\end{equation}
\begin{equation}
\label{Eq11} 
		m_{[ij]} = m_i + m_j - \frac{3b_{ij}}{4},
\end{equation}
where $b_{ij}$ represents the strong hyperfine interaction between quarks $i$ and $j$ with a strength given by Eq. \eqref{eq2}. 
These mass expressions include spin-dependent corrections arising from OGE, consistent with the QCD-inspired effective framework. Unlike Scenario I, the diquark masses of $D^{\{ij\}}$ and $D^{[ij]}$ in Scenario II exclude the multiplicative factor $3$ in the hyperfine term $b_{ij}$, avoiding the full absorption of hyperfine contributions into the diquark.\footnote{In the present construction, spin-independent color-Coulomb contributions are consistently included in the baseline OGE-based baryon mass framework~\cite{Karliner:2014gca}. Confinement-induced binding-energy effects are evaluated independently and added as explicit corrections. The effective diquark mass therefore reflects the internal short-distance structure fixed by baryon spectroscopy.} This separation enables the residual chromomagnetic interaction between the diquark and the third quark to be treated explicitly at the effective quark-diquark level. The hyperfine coupling between a diquark $D$ (composed of quarks $i$ and $j$) and the third quark $k$ is denoted $b_{\{ij\}k}$, with corresponding spin operators $\bm{s}_D$ and $\bm{s}_k$. As indicated by the color factor $\langle \vec{\lambda}_{\text{diquark}} \cdot \vec{\lambda}_k \rangle_{\text{singlet}}$ in Eq.~\eqref{casimir}, this interaction mirrors the color structure of the quark-antiquark coupling in mesons, allowing $b_{\{ij\}k}$ to be modeled analogously \cite{Pan:2023hwt, Chakrabarti:2010zz, Kiselev:2002iy, Kiselev:2001fw, Ebert:1996ec}. Within this construction, the short-distance chromomagnetic scale is derived directly from the OGE interaction, the same fundamental dynamics governing quark-quark hyperfine splittings. The contact matrix element, $\langle \psi | \delta^3(\vec{r}) | \psi \rangle$, isolates the resolved short-distance limit of the wavefunction \cite{DeRujula:1975qlm, Morpurgo:1989my}, leading to the hyperfine interaction coefficient:\begin{equation}\label{Eq6}b_{\{ij\}k} = \frac{32\pi\alpha_s}{9m_{\{ij\}}m_k} \langle\psi|\delta^3(\vec{r})|\psi\rangle.\end{equation}This hyperfine interaction coefficient naturally preserves the $1/(m_i m_j)$ inverse-mass hierarchy characteristic of OGE-driven interactions. By formulating this coupling strictly within quark-diquark degrees of freedom, we maintain the parameter-light nature of the framework without introducing an additional dynamical scale. Consequently, the general mass expressions for a baryon $B(ijk)$ can be parameterized through the effective masses of its constituent diquark and quark degrees of freedom as follows:
\begin{enumerate}	
    \item[1.] For octet baryons,
	\begin{enumerate}
		\item[(a)] Antisymmetric $\Lambda_{[i j]k}-$type ($J^P=\frac{1}{2}^+$) baryons take the form,
		\begin{equation}
			\begin{split}
				\label{Eq7}
				M_B &= m_{[ij]} + m_k\\
				&= m_{i} + m_{j} - \frac{3b_{ij}}{4} + m_k.              
			\end{split}
		\end{equation}
		\item[(b)] Symmetric $\Sigma_{\{i j\}k}-$type ($J^P=\frac{1}{2}^{\prime+}$) baryons manifest as,
		\begin{equation}
			\begin{split}
				\label{Eq8}
				M_{B^{\prime}}  &= m_{\{ij\}} 
				+ m_k - b_{\{ij\}k}\\
				&= m_{i} + m_{j} + \frac{b_{ij}}{4} + m_k - b_{\{ij\}k}.
			\end{split}
		\end{equation}
	\end{enumerate}
\item[2.] For decuplet ($J^P=\frac{3}{2}^+$) baryons,
    \begin{equation}
    \begin{split}
    \label{Eq9}
    M_{B^{*}}  &= m_{\{ij\}} + m_k + 
			b_{\{ij\}k}\\
		  &= m_{i} + m_{j} + \frac{b_{ij}}{4} + m_k + b_{\{ij\}k}.
    \end{split}
    \end{equation}
\end{enumerate}
It is worth to note that, only antisymmetric $\Lambda_{[i j]k}-$type baryons comprise of scalar diquark $D^{[ij]}$, whereas axial-vector diquark $D^{\{ij\}}$ is present in the symmetric $\Sigma_{\{i j\}k}-$type and decuplet baryons. Moreover, there is no spin interaction between a scalar diquark and the residual quark inside the baryon \cite{Jaffe:2004ph, Barabanov:2020jvn}. Therefore, the quark-diquark hyperfine interaction term $b_{\{ij\}k}$ always represents the interaction of an axial-vector diquark $D^{\{ij\}}$ with the residual quark $k$. 

In Scenario II, the constituent quark masses and strong hyperfine interaction terms ($m_i$, $b_{ij}$, and $b_{\{ij\}k}$) are obtained using the baryon mass relations formulated within the Scenario II framework and the experimentally known baryon masses from PDG \cite{ParticleDataGroup:2024cfk}. The parameter extraction relies on a careful selection of diquark configurations consistent with known experimental trends and lattice QCD insights, particularly favoring light-light diquarks in singly heavy baryons and heavy-heavy diquarks in doubly and triply heavy systems.

To reduce the parameter space, we adopt approximate flavor symmetry, with explicit breaking introduced through the constituent quark masses and hyperfine coefficients, supplemented by phenomenological diquark preferences primarily driven by chromomagnetic attraction, with subleading electromagnetic effects, favoring light flavor-antisymmetric configurations such as $ud$. These guiding principles were applied during the minimization procedure in the $SU(2)$ and $SU(3)$ sectors to identify the optimal diquark configurations with the lowest $\chi^2$ values. In contrast, no fitting was performed in the heavy flavor sectors; instead, input parameters were extracted directly from experimentally known heavy flavor baryon masses. In sectors with sparse experimental data, notably the bottom and multi-heavy baryons, hyperfine terms were calculated using approximate flavor symmetry relations. Importantly, unlike Scenario I, Scenario II does not include the multiplicative factor of $3$ in hyperfine interactions, thereby yielding a more transparent treatment of spin-dependent effects. Furthermore, our analysis reinforces the established trend that doubly heavy baryons are more favorably described by heavy-heavy diquarks, while singly heavy baryons prefer light-light configurations \cite{Chen:2014nyo, Ebert:2011kk, Ebert:2007nw, Faessler:2006ft, Albertus:2006ya, Ebert:2002ig, Ebert:1996ec}. By contrast, heavy-light diquarks emerge as less likely. Notably, restricting to these dominant configurations not only reflects the underlying dynamics but also reduces the number of free parameters, thereby improving the accuracy of mass predictions. For this reason, heavy-light diquark assumptions are omitted from our framework.\footnote{Heavy-light diquarks are not included; while relevant for exotics, their effects can be approximated via symmetry relations \cite{Mohan:2026blk}.} Within the QDEMF framework, Scenario II assigns light-light diquarks to singly heavy baryons and heavy-heavy diquarks to doubly heavy baryons. The diquark masses for all allowed flavor sectors are computed from the extracted parameters, as explained in Sec. \ref{Sec3}.

\subsection{Binding energy}
\label{BE}
Studies of doubly heavy baryon production in high energy $pp$, $ep$, $\gamma\gamma$, and $e^+e^-$ collisions show that heavy-diquark cross sections effectively bound the baryon yield \cite{Trunin:2016uks}. In these processes, two heavy quark-antiquark pairs are created and may recombine into a compact heavy diquark, which subsequently hadronizes into a doubly heavy baryon \cite{Zhan:2023jfm, Trunin:2016uks, Chen:2014frw}. The hadronization probability is then governed by the fragmentation of the heavy diquark, which is enhanced by the strong binding between heavy quarks \cite{Likhoded:2009zz}. As first established by Karliner and Rosner, this enhanced stability originates from an additional binding term arising from the short-range chromoelectric interaction between two heavy quarks, a spin-independent contribution that must be incorporated when more than one heavy quark is present in the hadron \cite{Karliner:2014gca}. This framework not only provides the theoretical foundation for heavy-diquark formation but also explains their experimental relevance in doubly heavy baryon production \cite{Karliner:2014gca, Zhang:2021yul}.\footnote{In singly heavy baryons, light-light diquarks act mainly as structural correlations. Lattice QCD \cite{Drenska:2010kg} supports attraction in scalar ($0^+$) diquarks, reflected in the $\Lambda_Q-\Sigma_Q$ mass splittings, though a universal binding term can not be cleanly isolated as in the heavy-heavy case.} The binding energy scales as $-\sum \alpha_s/r_{ij}$, where $r_{ij}$ is the interquark distance, and is significant only when both quarks are heavy and move nonrelativistically \cite{Zhang:2021yul}. Heavier quarks, with shorter Compton wavelengths, are more effectively confined within the interquark potential, implying that an additional binding term is essential for heavy-heavy and heavy-strange interactions to maintain consistency with light hadrons \cite{Karliner:2014gca,Zhang:2021yul}.  

The binding energy between quark pairs in a diquark can be inferred from that of quark-antiquark pairs in mesons \cite{Karliner:2014gca}. For a quark-antiquark pair $(Q\overline{Q}^{\prime})$, the binding energy is
\begin{equation}
\label{Eq13}
BE(Q\overline{Q}^{\prime}) = \frac{1}{4}(3M_V + M_P) - m_Q - m_{\overline{Q}^{\prime}},
\end{equation}
where $M_V$ and $M_P$ are the vector and pseudoscalar meson masses, and $m_Q$ ($m_{\overline{Q}^{\prime}}$) are the constituent quark (antiquark) masses for $Q = c, b$ and $Q^{\prime} = s, c, b$. The constituent quark masses are extracted from ground-state heavy-light mesons \cite{ParticleDataGroup:2024cfk}, assumed free from binding energy effects: $m_u = m_d = 305.821$ MeV (from $\rho^+$ and $\pi^+$), $m_s = 489.261$ MeV (from $K^{*+}$ and $K^+$), $m_c = 1665.530$ MeV (from $D^{*0}$ and $D^0$), and $m_b = 5007.594$ MeV (from $B^{*+}$ and $B^+$). Color $SU(3)$ considerations dictate that the interaction strength of a quark pair $QQ^{\prime}$ in an antitriplet ($\overline{3}_c$) is half that of $Q\overline{Q}^{\prime}$ in a singlet ($1_c$) \cite{Karliner:2014gca, DeRujula:1975qlm}. Hence, the diquark binding energy is
\begin{equation}
\label{Eq14}
BE(QQ^{\prime}) = \frac{BE(Q\overline{Q}^{\prime})}{2}.
\end{equation}
We calculate the binding energies of $Q\overline{Q}^{\prime}$ and $QQ^{\prime}$ pairs and incorporate them into mass predictions for $J^P = \frac{1}{2}^+$ and $J^P = \frac{3}{2}^+$ heavy baryons. The numerical results are discussed in the following section.
\section{Numerical results and discussions}
\label{Sec3}
In this work, we analyze diquark flavor combinations in baryons within the QDEMF using two formulations. Scenario I is based on the quark-quark interplay, where all three quark pairs contribute to the baryon mass through the full three-body wave function. Scenario II, by contrast, adopts the quark-diquark interplay, reducing the dynamics to an effective two-body system between a diquark and the remaining quark. In Scenario I, the baryon mass is described within the three-quark picture of Eq.~\eqref{eq1}, where all quark pairs contribute through their constituent masses $m_i$ and hyperfine couplings $b_{ij}$. The values of $m_i$ and $b_{ij}$, extracted from experimental baryon masses \cite{ParticleDataGroup:2024cfk} using the EMS and adopted from our earlier work \cite{Mohan:2022sxm}, are summarized in Table \ref{t1}. Although Eq.~\eqref{eq_diquark_pairs} recasts the system (Eq.~\eqref{eq1}) in a quark-diquark picture, the couplings $b_{ij}$ remain two-body quantities and thus preserve their physical meaning. Numerically, we find the $b_{ij}$ values to be unchanged under the change of picture across all baryon sectors. While our earlier study was limited to strange and charm baryons \cite{Mohan:2022sxm}, the present work extends the analysis to the bottom sector. Since experimental data for bottom baryons are limited, we estimate the hyperfine terms $b_{cb}$ and $b_{bb}$ using approximate flavor symmetry relations.  Subsequently, we use the spin-dependent expressions (Eqs. \eqref{axial} and \eqref{scalar}) along with these inputs to calculate the diquark masses in Scenario I, accounting the isospin symmetry breaking as in Ref. \cite{Mohan:2022sxm}. A detailed discussion on diquark masses is given in the subsection \ref{diq_mass}.

Furthermore, Scenario II provides a refined framework by explicitly incorporating both intra-diquark and quark-diquark hyperfine interactions. We independently determine the constituent quark masses $m_i$, intra-diquark hyperfine interaction terms $b_{ij}$, and quark-diquark interaction terms $b_{\{ij\}k}$ using baryon mass relations (Eqs. \eqref{Eq7}-\eqref{Eq9}) and experimental baryon masses \cite{ParticleDataGroup:2024cfk}. Using $N$ and $N^*$ baryon masses and their isospin splittings, we minimize the corresponding mass relations to obtain $m_u$, $m_d$, $b_{uu}$, $b_{ud}$, $b_{\{qq\}u}$, and $b_{\{qq\}d}$ in the $SU(2)$ flavor sector.\footnote{Note that we ignore the negligible impact of isospin breaking on $b_{\{ij\}k}$, and set $b_{\{uu\}u}=b_{\{ud\}u}=b_{\{dd\}u}$ to reduce parameters.} The minimization, performed using the MINUIT \cite{James:1975dr}, employs the $\chi^{2}$ function,
\begin{equation} 
\label{Eq12}
	\chi^{2} = \sum_{i} \bigg(\frac{M^{Theo}_{i} - M^{Expt}_{i}}{M^{Expt}_{i}}\bigg)^{2},
\end{equation}
where $M^{Theo}_{i}$ ($M^{Expt}_{i}$) are theoretical (experimental) masses given by Eq.~\eqref{Eq5} (PDG values \cite{ParticleDataGroup:2024cfk}). We obtain an excellent fit with $\chi^2\sim 0.0000075$. However, in $SU(2)$, the hyperfine interaction term $b_{dd}$ is computed directly from the mass of $\Delta^-$ baryon.\footnote{Since the $M_{\Delta^-}$ is not available experimentally, we use $M_{\Delta^-} = M_{\Delta^{++}} + 3(M_n - M_p)$ \cite{Rosner:1998zc}.}

Similarly, minimization of the $SU(3)$ baryon mass relations (Eqs.~\eqref{Eq7}–\eqref{Eq9}) against PDG inputs produces a fit of comparable accuracy ($\chi^2 \sim 0.000077$). A critical factor here is the choice of diquark flavor configuration. We found that configurations involving oppositely charged quarks ($ud$, $us$) are strongly favored, yielding superior fits in line with the mechanism of electromagnetic attraction. Conversely, identical-flavor diquarks produced unphysical parameters, such as a large negative $b_{ss}$. This preference for $ud$-type diquarks is corroborated by LQCD studies and deep-inelastic scattering results \cite{Drenska:2010kg, Jaffe:2004ph, Wilczek:2004im}.

In heavy flavor sectors, we use only dominant diquark configurations: light-light for singly heavy and heavy-heavy for multi-heavy baryons. This approach aligns with theory and reduces model complexity. Numerous studies \cite{Chen:2014nyo, Ebert:2011kk, Ebert:2007nw, Faessler:2006ft, Albertus:2006ya, Ebert:2002ig, Ebert:1996ec} support this choice, while heavy-light diquarks are excluded due to their tendency to yield unrealistically large hyperfine terms. We extract parameters directly from mass relations (Eqs. \eqref{Eq7}-\eqref{Eq9}) and experimental data \cite{ParticleDataGroup:2024cfk}. For bottom sector with scarce data, we estimate hyperfine terms via controlled $SU(3)$/$SU(4)$ flavor symmetry breaking (Eqs. \eqref{eq2}, \eqref{Eq6}, and Tables \ref{t1}-\ref{t2}). Finally, the diquark masses in both scenarios, using inputs from Tables \ref{t1} and \ref{t2}, are listed in Table \ref{t3}. In addition, the variation of the hyperfine mass splitting, $\Delta M = M_A - M_S$, with the corresponding diquark masses for scenarios~I and~II, is illustrated in Fig. \ref{Fig1}.

Unlike Scenario I, which includes all diquark configurations, Scenario II restricts heavy flavor diquarks to the three dominant cases, $D^{cc}$, $D^{cb}$, and $D^{bb}$, while omitting heavy-light diquarks. This selective treatment provides a consistent and tractable framework for conventional baryons and can be extended to exotic states. Using these inputs within the QDEMF, we compute the masses of heavy flavor baryons with $J^P=\tfrac{1}{2}^+$ and $\tfrac{3}{2}^+$ in both scenarios. To further refine predictions, we incorporate binding energy ($BE$) contributions, estimated from meson sector inputs (Table \ref{t4}). The correlated behavior of hyperfine splitting and binding energy across different diquark flavors is shown in Fig.~\ref{Fig2}. The final baryon masses, with and without $BE$ corrections, are presented in Tables \ref{t5}-\ref{t7}, alongside comparisons with PDG data \cite{ParticleDataGroup:2024cfk} and other theoretical models.\footnote{We list the obtained masses for light flavor baryons in Appendix \ref{A3}.} The detailed discussion of these results follows.

\subsection{Diquark masses in Scenario I and II} 
\label{diq_mass}
In this subsection, we present our diquark mass predictions, with Scenario~I results in column~3 and Scenario~II results in column~4 of Table \ref{t3}. Additionally, we provide a comparison with diquark mass predictions from other theoretical approaches. Note that we adopt standard conventions for baryon notation, where primed and unprimed labels distinguish whether the first-coupled quark pair forms a symmetric (spin-1) or antisymmetric (spin-0) configuration. For singly heavy baryons $(qqQ)$, the antisymmetric light-quark pair $[q_1q_2]$ ($S=0$) corresponds to $\Lambda_Q (\Xi_Q)$, while the symmetric pair $\{q_1q_2\}$ ($S=1$) corresponds to $\Sigma_Q (\Xi_Q^{\prime})$. In doubly heavy baryons $(qQQ)$, the heavy quarks can couple as an antisymmetric diquark $[QQ]$ ($S_D=0$) or a symmetric diquark $\{QQ\}$ ($S_D=1$). Standard conventions assign the unprimed baryon $|B\rangle$ to $\{QQ\}$ and the primed state $|B^{\prime}\rangle$ to $[QQ]$ \cite{Faessler:2009xn, Albertus:2006ya, Ebert:2002ig}. In our QDEMF, we adopt a unified scheme in which antisymmetric configurations are always unprimed and symmetric ones primed, for both singly and doubly heavy baryons.\footnote{Mixing between $|B\rangle$ and $|B^{\prime}\rangle$ states is neglected.} This notation is used consistently for both Scenario~I and II. We list our key observations as follows.

\begin{itemize}
    \item[(i)] The diquark masses in Scenario~I display a clear pattern governed by the hyperfine interaction. The splitting $\Delta M = M_A - M_S$ and ratio $R = M_A/M_S$ serve as clean probes of QCD spin dynamics, reflecting the strength of spin-spin correlations while largely canceling constituent mass effects. From Eqs.~\eqref{axial} and \eqref{scalar}, the axial-scalar mass difference $\Delta M = 3b_{ij}$ arises directly from the color-magnetic hyperfine term generated by OGE. The ratio $R$ (see the last row of Table \ref{t3}) provides a dimensionless measure of this effect, quantifying how the spin-spin interaction modifies the diquark’s effective binding energy relative to its spin-averaged mass, and thereby offering a scale-independent probe of correlation strength across flavor sectors. In this sense, $\Delta M$ and $R$ represent direct quantifications of diquark correlation strength and the corresponding binding energy induced by hyperfine dynamics \cite{Jaffe:2004ph}. 

    \item[(ii)] For light-light diquarks, especially the $ud$ pair, the scalar configuration experiences strong spin-spin attraction (see Eq.\eqref{scalar}), leading to a large binding energy and a very low mass ($M_S = 279.08$~MeV). This produces a sizable splitting $\Delta M \simeq 593$~MeV and ratio $R \simeq 3.13$, indicating deep binding and a pronounced dynamical preference for the flavor-antisymmetric, spin-0 scalar (``good”) diquark \cite{Jaffe:2004ph, Wilczek:2004im}. The $us$ and $ds$ diquarks follow the same trend with $\Delta M \sim 420$~MeV and $R \sim 1.74$, showing strong but somewhat weaker spin-spin correlations and binding. In this light-quark regime, the color-magnetic hyperfine interaction dominates, generating large mass splittings and substantial binding energies, underscoring the central role of light scalar diquarks as deeply bound components of baryons and multiquark states.

    \item[(iii)] As heavier quarks are introduced, the diquark mass splittings decrease systematically, consistent with the expected $1/(m_i m_j)$ scaling of the hyperfine interaction. In heavy-light diquarks such as $uc$ and $sc$, the splittings reduce substantially ($\Delta M \sim (120-140)$~MeV with ratios $R \sim 1.07$), reflecting the $1/m_Q$ suppression of color-magnetic effects, while for heavier partners like $ub$ or $sb$, the ratio approaches unity ($R \approx 1.01$), indicating near-decoupling of the heavy-quark spin. The trend, clearly illustrated in Fig.~\ref{Fig1}, shows an inverse correlation between the axial-scalar mass splitting $\Delta M = M_A - M_S$ and the diquark mass, directly tracing the weakening of the spin-spin (hyperfine interaction) term. This systematic suppression signifies the gradual onset of HQSS, wherein the heavy-quark spin decouples from the light degrees of freedom, rendering scalar and axial-vector diquark states nearly degenerate. The convergence of the curves at higher diquark masses in Fig.~\ref{Fig1} corroborates this behavior, reflecting the diminishing role of the color-magnetic hyperfine interaction and the transition toward a spin-independent, compact diquark structure, as exemplified by the $cb$ diquark exhibiting a small splitting of $\Delta M \sim 15$~MeV with $R \simeq 1$. In this regime, QCD dynamics are effectively spin-independent, and the diquark behaves as an almost point-like constituent, in accordance with HQSS \cite{Albertus:2006ya, Jenkins:1992nb}.

    \item[(iv)] Motivated by HQSS considerations, we identify effective diquark configurations in Scenario II. In this framework, the light-light diquarks dominate singly heavy baryons due to their strong scalar attraction, while heavy-heavy diquarks form the preferred building blocks of doubly and triply heavy baryons, stabilized by color-Coulomb binding and spin-independence implied by HQSS. For light diquarks, Scenario II yields systematically smaller splittings than Scenario I because only the intra-diquark hyperfine term ($b_{ij}$) contributes. For example, the $ud$ pair shows $\Delta M = b_{ud} \sim 107$ MeV with $R \simeq 1.17$, compared to the much larger splitting in Scenario I. A similar pattern is observed for the $us$ and $ds$ diquarks,\footnote{Although the $b_{ij}$ values differ by $26$ MeV for the $us$ and $ds$ pairs, we have listed the values true to the fit. However, this difference induces a mild deviation in the resulting diquark masses.} demonstrating that while light scalar diquarks remain strongly correlated in both pictures, Scenario II reflects a weaker overall attraction. In the heavy sector, the $cb$ diquark displays a very small splitting of $\Delta M \sim 4.5$ MeV with $R \simeq 1$, fully consistent with HQSS predictions of near-degeneracy between axial-vector and scalar configurations. On the other hand, heavy-light diquarks are excluded because they neither maximize hyperfine attraction (as in light-light systems) nor benefit from HQSS-driven degeneracy (as in heavy-heavy systems), making them dynamically less stable within baryons. Scenario II therefore provides a more refined and explicit estimate of diquark masses, whereas Scenario I employs a more inclusive treatment of hyperfine interactions, effectively absorbing contributions from all three quark-quark pairs, an approach that may be regarded as a broader estimate of the hyperfine splitting. Despite these structural differences, both scenarios yield consistent qualitative conclusions: light scalar diquarks show strong correlations, while heavy diquarks behave nearly spin-independently. 

    \item[(v)] A comparison with other theoretical approaches, including PM \cite{Ferretti:2019zyh}, CIM \cite{Yin:2019bxe}, BSE \cite{MoosaviNejad:2020nsl}, $\chi$QM~\cite{Kim:2021ywp}, RQDM \cite{Santopinto:2014opa}, and RQM \cite{Ebert:2002ig, Ebert:2011kk}, shows both numerical consistency and characteristic trends. For light diquarks, these models predict scalar masses in the $(600-950)$~MeV range and axial-vector masses around $(840-1300)$~MeV, giving ratios $R \sim (1.1-1.6)$, in agreement with our Scenario~II values. In the heavy-light sector, our Scenario~I results and those of PM \cite{Ferretti:2019zyh}, CIM \cite{Yin:2019bxe}, and BSE \cite{MoosaviNejad:2020nsl} exhibit smaller splittings and ratios closer to unity, reflecting the expected suppression of hyperfine effects. For heavy-heavy diquarks, PM \cite{Ferretti:2019zyh}, CIM \cite{Yin:2019bxe}, and RQM \cite{Ebert:2002ig} predict near-degenerate axial-vector and scalar masses, consistent with our results and HQSS expectations. These comparisons confirm that the patterns observed in our calculations, strong spin-spin correlations in light diquarks, suppressed splittings in heavy-light systems, and near-degeneracy in heavy-heavy diquarks, are robust across different theoretical frameworks.
\end{itemize}
It may be emphasized that, complementing the inclusive treatment of Scenario~I, Scenario~II provides physically refined diquark masses, respecting HQSS and yielding realistic correlations and binding energies across both light and heavy sectors. We now proceed with the baryon mass analysis in the following subsection using these diquark masses in both scenarios.

\subsection{$J^P=\frac{1}{2}^+$ and $J^P=\frac{3}{2}^+$ heavy baryon masses}\label{mass_discussion}
We present our predictions for the $J^P=\tfrac{1}{2}^+$ and $J^P=\tfrac{3}{2}^+$ heavy baryon masses in Tables \ref{t5}-\ref{t7} for both Scenario I and Scenario II. The corresponding results incorporating the binding energy corrections from Table \ref{t4} are also listed. We further compare our predictions with available experimental data \cite{ParticleDataGroup:2024cfk} and with results from various theoretical models. We list out main observations as follows.
\begin{itemize}
\item[(i)] Our predictions for the singly charmed baryon masses show excellent agreement with the experimental values \cite{ParticleDataGroup:2024cfk} in Scenario II, with an average deviation of less than $1\%$, except for the $\Sigma_c^{(*)}$ states. For the $\Sigma_c^{(*)}$ baryons, the predicted masses remain consistent with the PDG data, with a maximum deviation of $\sim\mathcal{O}(3\%)$. We emphasize that the results for singly heavy baryons remain unaffected by the binding energy terms in Scenario II, as these systems involve lighter diquarks. As discussed earlier, the binding energy contributions are defined only for interactions between heavy-heavy ($c-b$) and heavy-strange ($c-s$ or $b-s$) quark pairs~\cite{Karliner:2014gca}, and therefore do not influence the singly heavy baryon sector. Consequently, the choice of a light-light ($u ~\text{or}~ d$) diquark configuration in singly heavy baryons makes the Scenario II masses independent of binding energy contributions. 

\item[(ii)] In Scenario I, our predictions also agree well with the PDG values \cite{ParticleDataGroup:2024cfk} when excluding the binding energy terms. Specifically, the Scenario I predictions obtained without these terms coincide numerically with our earlier results reported in Ref. \cite{Mohan:2022sxm} (see Table IV), since both frameworks employ identical input parameters.  Similar to Scenario II, the $\Sigma_c^{(*)}$ baryon masses exhibit a maximum deviation of $\sim\mathcal{O}(3\%)$. Crucially, in Scenario I we incorporate pairwise binding energy terms to more comprehensively account for quark-quark interactions. Accordingly, all binding energy terms listed in column 5 of Table \ref{t4} are included in the Scenario I calculations. As discussed earlier within the QDEMF, constituent quark masses inherently incorporate the spin-independent confinement and color-Coulomb contributions, while the hyperfine interactions are treated explicitly. The phenomenological binding energy term introduced here quantifies intra-diquark correlations in the color-antisymmetric $\overline{3}_c$ channel. Calibrated from mesonic systems and scaled by the corresponding color factor,\footnote{For $\overline{3}_c$ diquarks, the color factor $C_{qq} = \tfrac{1}{2}C_{q\overline{q}}$, consistent with OGE.} this term represents the short-range color-Coulomb enhancement beyond the mean-field effects embedded in the constituent masses. Upon their inclusion, the Scenario I mass predictions for singly charmed baryons ($\Xi_c^{(\prime)}$, $\Xi_c^*$, and $\Omega_c^{(*)}$) deviate roughly by $3\%$ with respect to the PDG values \cite{ParticleDataGroup:2024cfk}. This systematic downward shift arises because these terms enter the mass relation as negative corrections, representing the attractive inter-quark potential. This destructive interference lowers the predicted total baryon mass by a few percent. In this sense, while the downward mass shift signifies an overall attractive interaction among quarks, it is difficult to interpret it as a well-defined diquark binding energy, since Scenario I does not assign a concrete spin-flavor structure to any particular quark pair.\footnote{The smaller $m_c$ and larger effective $b_{ij}$ values in Scenario I, together with the inclusion of $BE$ estimates, effectively mimic enhanced quark-quark correlations in the absence of specific diquark state, yielding trends comparable to Scenario II.} 

\item[(iii)]  As we progress from singly heavy to doubly and triply heavy baryons, a clear systematic trend emerges. Fig. \ref{Fig2} illustrates that for heavy diquarks, the binding energy becomes increasingly negative with rising diquark mass, from $cc$ to $cb$ and further to $bb$, while the hyperfine splitting between the axial-vector and scalar configurations decreases, consistent with $1/(m_i m_j)$ suppression of the chromomagnetic interaction. This inverse correlation reflects the HQSS behavior, wherein hyperfine effects diminish in the heavy-quark limit, and color-Coulomb attraction increasingly dominates within compact heavy diquarks. Consequently, inclusion of the binding energy term systematically lowers the predicted masses of baryons containing multiple heavy quarks, shifting them slightly below the corresponding experimental and LQCD values. This downward shift is more pronounced in Scenario I, whereas in Scenario II the inclusion of binding energy improves agreement with data, particularly for the doubly and triply charmed systems discussed next. This consistent trend across the heavy-quark spectrum confirms that binding energy corrections are essential to capture intra-diquark dynamics and ensure physically grounded mass predictions for baryonic and exotic states.

\item[(iv)] In the doubly and triply charmed sector, our Scenario II predictions without binding energy corrections already agree well with PDG \cite{ParticleDataGroup:2024cfk, LHCb:2021eaf} and LQCD \cite{Bahtiyar:2022nqw, Bahtiyar:2020uuj} results for the $\Xi_{cc}^{(*)}$ states, while the $\Omega_{cc}^{(*)+}$ and $\Omega_{ccc}^{*++}$ masses are overestimated by $\sim (3-5)\%$, consistent with HQSS expectations. Including binding energy correction leads to a systematic reduction in the predicted masses, bringing the Scenario II results for $\Omega_{cc}^{(*)+}$ into excellent agreement with recent LQCD results \cite{Bahtiyar:2022nqw, Bahtiyar:2020uuj, Mathur:2018rwu}, while $\Xi_{cc}^{(*)}$ and $\Omega_{ccc}^{*++}$ agree within $\sim2\%$. Interestingly, the $BE$ term inverts the $1.4\%$ overestimation of $\Xi_{cc}$ masses to a $2\%$ underestimation, while the magnitude of deviation remains nearly constant. In contrast, Scenario I predictions for $\Xi_{cc}^{++}$ and $\Xi_{cc}^{+}$ remain in good agreement with both experimental \cite{ParticleDataGroup:2024cfk, LHCb:2021eaf} and LQCD \cite{Bahtiyar:2022nqw} data, while $\Xi_{cc}^*$ and $\Omega_{cc}^{(*)+}$ are consistent with lattice results \cite{Bahtiyar:2022nqw, Bahtiyar:2020uuj, Mathur:2018rwu} even without the $BE$ term. A moderate deviation of about $3\%$ persists for the $\Omega_{ccc}^{*++}$ in Scenario I. As in the singly charmed case, incorporating pairwise $BE$ terms in Scenario I further lowers the masses of doubly and triply charmed baryons, thereby increasing their deviation from PDG and LQCD values to roughly $(3-5)\%$. Overall, Scenario II offers a balanced treatment of hyperfine and binding effects, consistent with HQSS, and achieves superior agreement with experiment and LQCD. Its success further supports the effective quark-diquark picture, where compact color-antisymmetric correlations act as diquark substructures, reinforcing the physical relevance of including an intra-diquark binding term within QDEMF \cite{Karliner:2014gca, Roberts:2007ni, Ebert:2002ig}.

\item[(v)] We observe that the consistency of our mass predictions with the PDG values improves further in the case of bottom baryons. In Scenario II, the predicted masses of singly bottom baryons exhibit excellent agreement with the PDG data \cite{ParticleDataGroup:2024cfk}, remaining within $0.4\%$, except for the $\Sigma_b^{(*)}$ states. The $\Sigma_b^{(*)}$ baryon masses also show reasonable consistency, deviating only by $1.4\%$ from the PDG values \cite{ParticleDataGroup:2024cfk}. As discussed previously for the charm sector, the Scenario II results for singly heavy baryons are unaffected by the $BE$ term. Furthermore, in Scenario I, all singly bottom baryon mass predictions are in very good agreement with experimental measurements, except for the $\Omega_b^{(*)-}$, whose masses are overestimated by $1.3\%(2\%)$, compared to PDG (LQCD) results \cite{ParticleDataGroup:2024cfk, Mohanta:2019mxo}. Notably, the predicted $\Omega_b^{(*)-}$ masses show improved agreement with PDG and LQCD values upon inclusion of the pairwise $BE$ terms, which are relevant for diquarks with strange content. Interestingly, we observe an opposite trend to the singly charmed baryons: the extracted bottom-quark mass, $m_b$, is larger in Scenario I than in Scenario II, while the effective hyperfine terms remain nearly identical and small in both cases. The suppression of spin-dependent effects in the heavy-quark sector, consistent with HQSS expectations, causes the heavier $m_b$ in Scenario I to dominate the results. The modest overestimation introduced thereby is effectively compensated by the inclusion of binding energy corrections.

\item[(vi)] As we progress toward heavier systems, Scenario~II distinctly establishes itself as the more physically consistent and predictive framework compared to Scenario I. Its formulation, grounded in an effective diquark configuration, inherently embodies the hierarchical suppression of spin-dependent interactions, in excellent accordance with the expectations of HQSS. It is important to note that only a few experimentally measured bottom baryon masses are available to constrain the hyperfine interaction terms. This limitation poses a challenge, which we address by invoking approximate symmetry relations. It should be emphasized, however, that symmetry breaking is inherently incorporated into our formalism, as distinct quark masses and $b_{ij}$ parameters are employed within these relations (given in Tables \ref{t1} and \ref{t2}) to compensate for symmetry-breaking effects. Consequently, most of our baryon masses are pure theoretical predictions that are in excellent agreement with the experimentally measured masses \cite{ParticleDataGroup:2024cfk}. Therefore, QDEMF establishes a robust framework that reliably predicts consistent results with experimental data in both pictures.
		
\item[(vii)] Our Scenario~II results for doubly heavy (charmed bottom and doubly bottom) baryons are initially overestimated by about $\mathcal{O}(3\%)$ when binding energy effects are excluded. Including the binding energy terms brings the predictions into excellent agreement with LQCD results \cite{Mohanta:2019mxo, Mathur:2018epb}, reducing deviations to below $1\%$. This confirms that binding energy contributions are essential for systems containing multiple heavy quarks. A notable feature arises in the $\Xi_{cb}^{(\prime)}$ and $\Omega_{cb}^{(\prime)}$ spectra: in Scenario~I, baryons with scalar diquarks lie below their axial-vector counterparts, whereas Scenario~II reverses this ordering, consistent with earlier studies \cite{Farhadi:2023ucs, Faessler:2009xn, Albertus:2006ya, Ebert:2002ig}, such that $M(\Xi_{cb}) > M(\Xi_{cb}^{\prime})$ and $M(\Omega_{cb}) > M(\Omega_{cb}^{\prime})$. Experimental verification of these mass hierarchies would provide an excellent test of the underlying theoretical framework.

\item[(viii)] For triply heavy baryons, our Scenario II mass predictions (with binding energy terms) are overestimated by $(2-5)\%$ relative to the LQCD results of Ref. \cite{Mohanta:2019mxo}, but align closely with LQCD \cite{Mathur:2018epb}, showing average deviations of $\mathcal{O}(2\%)$. Inclusion of pairwise binding energy terms in Scenario~I effectively reduces the deviations from about $7\%$ to below $1\%$, achieving excellent consistency with LQCD \cite{Mohanta:2019mxo, Mathur:2018epb}. In summary, while the pairwise binding energy corrections in Scenario~I yield a coherent description of multi-heavy systems, Scenario~II by emphasizing dominant diquark configurations, provides a more compact, physically grounded, and predictive framework. Its results demonstrate remarkable consistency with both experimental and LQCD data, establishing Scenario~II as the more realistic representation of diquark dynamics in heavy baryons. 
\end{itemize}

Our predictions for charm and bottom baryon masses show strong agreement with a broad range of theoretical approaches, including BSE \cite{Farhadi:2023ucs}, $\chi$QM \cite{Kim:2021ywp}, RQM \cite{Ebert:2011kk}, HB$\chi$PT \cite{Jiang:2014ena, Yao:2018ifh}, HCQM \cite{Shah:2016mig, Shah:2016vmd, Shah:2017jkr, Shah:2017liu}, BM \cite{Zhang:2021yul}, QCDSR \cite{Liu:2007fg, Aliev:2012iv, Aliev:2012tt}, and NRQM \cite{Ortiz-Pacheco:2023kjn}. Effective field theories such as HB$\chi$PT \cite{Jiang:2014ena, Yao:2018ifh} expand baryon masses systematically in light meson momentum and inverse heavy baryon mass using an effective Lagrangian consistent with QCD symmetries. Quark potential models like the HCQM \cite{Shah:2016mig, Shah:2016vmd, Shah:2017jkr, Shah:2017liu} and harmonic oscillator frameworks \cite{Ortiz-Pacheco:2023kjn} treat baryons as three-body bound systems governed by confining color-Coulomb and hyperfine interactions, achieving good consistency through optimized parameters. The BM \cite{Zhang:2021yul} confines relativistic quarks within a finite bag, incorporating chromomagnetic and empirical heavy quark binding corrections, though its isospin symmetry assumption restricts predictions to specific multiplets. In contrast, momentum-independent approaches such as the CIM \cite{Yin:2019bxe} and simplified BSE models \cite{Farhadi:2023ucs} often overestimate or locally deviate, for example, $\Lambda_c^+$ is overestimated by $\sim15\%$ and $\Xi_c^{*0}$ underestimated by $\sim7\%$, mainly due to the absence of dynamical momentum-dependence and simplified diquark spin couplings.

In the bottom sector, the BSE \cite{Farhadi:2023ucs}, $\chi$QM \cite{Kim:2021ywp}, RQM \cite{Ebert:2011kk}, BM \cite{Zhang:2021yul}, QCDSR \cite{Liu:2007fg}, and NRQM~\cite{Ortiz-Pacheco:2023kjn} models reproduce experimental masses reasonably well \cite{ParticleDataGroup:2024cfk}, though most rely on parameter fitting in the heavy flavor sector. For instance, $\chi$QM \cite{Kim:2021ywp} and NRQM \cite{Ortiz-Pacheco:2023kjn} tune their parameters through simultaneous fits to experimental and LQCD baryon masses, while QCDSR \cite{Aliev:2012iv, Aliev:2012tt} and NRQM \cite{Roberts:2007ni} frameworks typically show larger deviations up to $\sim6\%$, for states such as $\Xi_{cb}^{*+}$, $\Omega_{cb}^{*0}$, and triply heavy baryons. In contrast, the QDEMF determines constituent quark masses and hyperfine terms directly from precise experimental inputs \cite{ParticleDataGroup:2024cfk} without any parameter fitting, offering a parameter-independent and self-consistent description of singly, doubly, and triply heavy baryons. Our predictions remain within $(1-2)\%$ of experimental \cite{ParticleDataGroup:2024cfk} and LQCD \cite{Mohanta:2019mxo, Mathur:2018epb} values, while other models typically deviate by $(2-6)\%$.

Overall, both scenarios~I and~II of the QDEMF reproduce the charm and bottom baryon spectra with excellent precision. Scenario~I, incorporating pairwise binding energy effects, provides a consistent description for bottom and multi-heavy systems, whereas Scenario~II, emphasizing effective heavy-heavy diquark configurations, naturally captures HQSS-driven mass hierarchies. Together, they offer a comprehensive, parameter-free framework that unifies the heavy baryon spectrum across flavor sectors.

\section{Summary and Conclusions}
\label{Sec4}
In this work, we present a novel QDEMF formulated on the effective masses of quarks and diquarks inside baryons. The effective masses are evaluated using quark-quark interactions mediated by OGE, and two complementary scenarios are explored within this framework. Scenario I, the quark-quark interaction picture, considers all possible diquark combinations among the three constituent quarks. In contrast, Scenario II, the quark-diquark interaction picture, employs specific diquark configurations with fixed spin-flavor degrees of freedom to effectively describe the baryon system. Within this framework, we predicted the masses of heavy flavor baryons with $J^P=\frac{1}{2}^+$ and $J^P=\frac{3}{2}^+$, identified the corresponding effective diquarks, and estimated their masses, mass splittings, and ratios in both scenarios. Crucially, we incorporated the binding energy contributions between heavy-heavy ($cc, cb, bb$) and heavy-strange ($cs, bs$) quark pairs, pairwise in Scenario I and through the specific (intrinsic) diquark configuration in Scenario II. The baryon masses obtained from both scenarios show excellent agreement with experimental results reported by the PDG \cite{ParticleDataGroup:2024cfk}, LQCD calculations \cite{Bahtiyar:2020uuj, Mohanta:2019mxo, Mathur:2018rwu, Mathur:2018epb}, and other theoretical models. These findings demonstrate that the proposed QDEMF provides a consistent and reliable description of baryon properties within the quark-diquark picture and could serve as a promising framework for understanding exotic states~\cite{Mohan:2026blk}. Our major conclusions are summarized as follows:

\begin{itemize}

\item Scenario I, incorporating all quark-quark hyperfine interactions, represents a broader aspect of spin-spin correlations, while Scenario II isolates the dominant intra-diquark effects, offering a more physically transparent effective description. Together, they capture the transition from strong spin coupling in light diquarks to HQSS-driven spin independence in heavy diquarks, outlining a coherent, flavor-dependent picture of QCD spin dynamics across the low-lying heavy baryon spectrum.
Light scalar diquarks exhibit strong color-magnetic binding and large hyperfine splittings that scale approximately as $1/(m_i m_j)$, progressively weakening in heavy-light systems and approaching degeneracy in the heavy-quark limit, consistent with HQSS expectations. This behavior constitutes a quantitative realization of emergent HQSS within the QDEMF. Notably, the scaling pattern is reproduced consistently across both scenarios, indicating that the emergence of heavy-quark symmetry is a robust consequence of the underlying chromomagnetic dynamics rather than a model-dependent artifact.

\item The diquark flavor configuration critically influences Scenario II mass predictions. Oppositely charged pairs such as $ud$ and $us$ are energetically favored due to electromagnetic attraction. Furthermore, these diquarks dominate singly heavy baryons, while heavy-heavy diquarks form the core of doubly and triply heavy systems. This hierarchy is consistent with HQSS expectations and aligns with lattice and model-based studies of heavy-baryon structure.

\item In both scenarios, inclusion of binding energy corrections proves essential for realistic heavy-baryon mass predictions. Scenario I captures the cumulative effect of pairwise quark interactions, while Scenario II integrates these contributions through the intrinsic diquark structure, achieving closer agreement with experiment and LQCD. Together, these results identify binding energies as a key dynamical ingredient in achieving quantitatively accurate baryon masses, and suggest that analogous contributions are likely to play a similarly important role in more complex multiquark systems.

\item We conclude that incorporating a mass-dependent binding energy is essential for a physically accurate description of heavy diquarks. This term quantitatively captures the transition from chromomagnetic to color-Coulomb dominance, a defining feature of HQSS. Its inclusion in Scenario I provides a broader, comprehensive estimate, whereas Scenario II offers a more refined and physically transparent picture, achieving superior agreement with experimental and lattice results. The quantitative results support this interpretation: for charmed baryons, Scenario I reproduces the masses well without the binding term, while Scenario II, with binding energy included, aligns more closely with experimental \cite{ParticleDataGroup:2024cfk, LHCb:2021eaf} and LQCD \cite{Bahtiyar:2022nqw, Bahtiyar:2020uuj, Mathur:2018rwu} data, maintaining deviations within $2\%$ across the sector.

\item In the bottom sector, inclusion of binding energy in both scenarios ensures excellent consistency with available data, with Scenario I slightly overestimating and  Scenario II achieves the most accurate overall agreement with LQCD \cite{Mohanta:2019mxo, Mathur:2018epb}, particularly for doubly and triply heavy systems.

\end{itemize} 
The QDEMF establishes a predictive and internally consistent framework for heavy baryon spectroscopy. Scenario II, in particular, provides a robust and efficient quark-diquark description that captures the observed mass hierarchies while maintaining strong agreement with experimental and lattice data using a minimal and parameter-independent set of inputs. By dynamically mapping the three-body problem onto an effective two-body description, the formalism achieves both computational simplicity and physical transparency while preserving heavy-quark symmetry trends. The calibrated diquark masses and chromomagnetic couplings extracted here therefore define robust effective degrees of freedom for heavy-hadron structure. In this sense, the present baryon-level calibration provides a controlled foundation for systematically extending quark-diquark dynamics to more complex multiquark systems.

\section*{Acknowledgment}
The author RD gratefully acknowledges the financial support by the Department of Science and Technology (SERB:TAR/2022/000606), New Delhi. Part of this work was carried out under the earlier project (SERB:CRG/2018/002796), whose support is also gratefully acknowledged.
\newpage
\appendix
\section*{Appendix}
\section{$SU(6)$ diquark wave functions}
\label{A1}
The axial-vector diquarks $A_1^+$ and $A_1^0$ denote the spin projections $S_z = +1$ and $S_z = 0$, respectively, for the $uu$ flavor configuration. The scalar diquark $S_1$ represents the $ud$ flavor configuration with spin projection $S_z = 0$.
\begin{table}[h!]
	\begin{tabular}{l l}\hline
		\multicolumn{2}{l}{($6_f \otimes 3_s)\subset21$}\\ \hline
		\multicolumn{2}{l}{}\\
		$A_1^+ = |u\uparrow u\uparrow \rangle$ & ~~~$A_1^0 = \frac{1}{\sqrt{2}} |u\uparrow u\downarrow + u\downarrow u\uparrow \rangle$\\
		$A_2^+ = \frac{1}{\sqrt{2}} |u\uparrow d\uparrow + d\uparrow u\uparrow \rangle$ & ~~~$A_2^0 = \frac{1}{2} |u\uparrow d\downarrow + u\downarrow d\uparrow + d\uparrow u\downarrow + d\downarrow u\uparrow \rangle$\\
		$A_3^+ = |d\uparrow d\uparrow \rangle$ & ~~~$A_3^0 = \frac{1}{\sqrt{2}} |d\uparrow d\downarrow + d\downarrow d\uparrow \rangle$\\
		$A_4^+ = \frac{1}{\sqrt{2}} |u\uparrow s\uparrow + s\uparrow u\uparrow \rangle$ & ~~~$A_4^0 = \frac{1}{2} |u\uparrow s\downarrow + u\downarrow s\uparrow + s\uparrow u\downarrow + s\downarrow u\uparrow \rangle$\\
		$A_5^+ = \frac{1}{\sqrt{2}} |d\uparrow s\uparrow + s\uparrow d\uparrow \rangle$ & ~~~$A_5^0 = \frac{1}{2} |d\uparrow s\downarrow + d\downarrow s\uparrow + s\uparrow d\downarrow + s\downarrow d\uparrow \rangle$\\
		$A_6^+ = |s\uparrow s\uparrow \rangle$ & ~~~$A_6^0 = \frac{1}{\sqrt{2}} |s\uparrow s\downarrow + s\downarrow s\uparrow \rangle$\\
		\multicolumn{2}{l}{}\\ \hline
		\multicolumn{2}{l}{$(\overline{3}_f \otimes 1_s)\subset21$}\\ \hline
		\multicolumn{2}{l}{}\\
		$S_1 = \frac{1}{2} |u\uparrow d\downarrow - u\downarrow d\uparrow - d\uparrow u\downarrow + d\downarrow u\uparrow \rangle$\\
		$S_2 = \frac{1}{2} |u\uparrow s\downarrow - u\downarrow s\uparrow - s\uparrow u\downarrow + s\downarrow u\uparrow \rangle$\\
		$S_3 = \frac{1}{2} |d\uparrow s\downarrow - d\downarrow s\uparrow - s\uparrow d\downarrow + s\downarrow d\uparrow \rangle$\\ 
		\multicolumn{2}{l}{}\\ \hline
	\end{tabular}
\end{table}
\section{Quark-diquark wave functions of $J^P = \frac{1}{2}^+$ and $J^P = \frac{3}{2}^+$ $SU(3)$ baryons}
\label{A2}
In the case of $J^P = \frac{1}{2}^+$ baryons, the axial-vector and scalar diquarks equally contribute to the total baryon wave function.
\begin{itemize}
	\item [(I)] For $(iik)-$type $J^P = \frac{1}{2}^+$ baryons,
\begin{equation*}
	|p \uparrow\rangle = \frac{1}{\sqrt{18}}\big|2A_1^+d\downarrow - \sqrt{2}A_1^0d\uparrow + A_2^0u\uparrow -\sqrt{2}A_2^+u\downarrow + 3S_1u\uparrow \big\rangle, 
\end{equation*}
\item [(II)] For $(ijk)~\Lambda-$type baryons,
\begin{equation*}
	|\Lambda^0 \uparrow\rangle = \frac{1}{\sqrt{12}}\big|\sqrt{2}A_4^+d\downarrow - A_4^0d\uparrow + A_5^0u\uparrow -\sqrt{2}A_5^+u\downarrow + 2S_1s\uparrow + S_2d\uparrow - S_3u\uparrow \big\rangle.
\end{equation*}
\item [(III)] For $(ijk)~\Sigma-$type baryons,
\begin{equation*}
	|\Sigma^0 \uparrow\rangle = \frac{1}{6}\big|2\sqrt{2}A_2^+s\downarrow - 2A_2^0s\uparrow + A_4^0d\uparrow -\sqrt{2}A_4^+d\downarrow + A_5^0u\uparrow -\sqrt{2}A_5^+u\downarrow + 3S_2d\uparrow + 3S_3u\uparrow \big\rangle.
\end{equation*}
\item [(IV)] In the case of  $J^P = \frac{3}{2}^+$ baryons, only axial-vector diquarks contribute to the total wave function.  
\begin{itemize}
	\item [(a)] For $(iii)-$type baryons,
\begin{equation*}
	|\Delta^{++} \uparrow\rangle = \big|A_1^+u\uparrow \big\rangle, 
\end{equation*}

\item [(b)] For $(iik)-$type baryons,
\begin{equation*}
	|\Delta^+ \uparrow\rangle = \frac{1}{\sqrt{3}}\big|A_1^+d\uparrow + \sqrt{2}A_2^+u\uparrow \big\rangle, 
\end{equation*}
\item [(c)] For $(ijk)-$type baryons,
\begin{equation*}
	|\Sigma^{*0} \uparrow\rangle = \frac{1}{\sqrt{3}}\big|A_2^+s\uparrow + A_4^+d\uparrow + A_5^+u\uparrow \big\rangle.
\end{equation*}
\end{itemize}
\end{itemize}
\section{Mass expressions of $J^P = \frac{1}{2}^+$ and $J^P = \frac{3}{2}^+$ $SU(3)$ baryons}
\label{A3}
The mass expressions of $SU(3)$ baryons in the quark-quark interaction picture (Scenario I) of QDEMF are obtained from the aforementioned quark-diquark wave functions of baryons. These expressions are obtained in terms of the masses of quarks and diquarks. For $J^P = \frac{1}{2}^+$ baryons, both axial-vector and scalar diquarks contribute equally to the total mass.
\begin{itemize}
	\item [(I)] For $(iik)-$type $J^P = \frac{1}{2}^+$ baryons,
\begin{equation*}
	M_p = \frac{1}{3}(m_{A_1} + m_d) + \frac{1}{6}(m_{A_2} + m_u) + \frac{1}{2}(m_{S_1} + m_u),
\end{equation*}
\item [(II)] For $(ijk)~\Lambda-$type baryons,
\begin{equation*}
	M_{\Lambda^0} = \frac{1}{4}(m_{A_4} + m_d) + \frac{1}{4}(m_{A_5} + m_u) + \frac{1}{3}(m_{S_1} + m_s) + \frac{1}{12}(m_{S_2} + m_d) + \frac{1}{12}(m_{S_3} + m_u).
\end{equation*}
\item [(III)] For $(ijk)~\Sigma-$type baryons,
\begin{equation*}
	M_{\Sigma^0} = \frac{1}{3}(m_{A_2} + m_s) + \frac{1}{12}(m_{A_4} + m_d) + \frac{1}{12}(m_{A_5} + m_u) + \frac{1}{4}(m_{S_2} + m_d) + \frac{1}{4}(m_{S_3} + m_u).
\end{equation*}
\item [(IV)] In the case of  $J^P = \frac{3}{2}^+$ baryons, the total mass is determined solely by the contribution of axial-vector diquarks and the residual quarks.  
\begin{itemize}
\item [(a)] For $(iii)-$type baryons,
\begin{equation*}
	M_{\Delta^{++}} = m_{A_1} + m_u.
\end{equation*}
\item [(b)] For $(iik)-$type baryons,
\begin{equation*}
	M_{\Delta^{+}} = \frac{1}{3}(m_{A_1} + m_d) + \frac{2}{3}(m_{A_2} + m_u). 
\end{equation*}

\item [(c)] For $(ijk)-$type baryons,
\begin{equation*}
	M_{\Sigma^{*0}} = \frac{1}{3}(m_{A_2} + m_s) + \frac{1}{3}(m_{A_4} + m_d) + \frac{1}{3}(m_{A_5} + m_u). 
\end{equation*}
\end{itemize}
\end{itemize}
Proceeding in a similar way, the wave functions and mass expressions of baryons up to $SU(5)$ flavor sector can be formulated based on their quark flavor content.
\newpage
\noindent\textbf{Numerical values of light baryon masses:} The masses of the light baryons, computed under Scenario I and Scenario II, are summarized in Table \ref{mass_light}.
\begin{table}[ht]
		\centering
		\captionof{table}{Masses of light baryons (in MeV).} 
		\label{mass_light}
		\begin{tabular}{|c|c|c|c|c|}	\hline  
			 \multicolumn{2}{|c|}{Baryons\footnotemark[1]} & Sc. II & Sc. I \cite{Mohan:2022sxm} & PDG\footnotemark[4] \cite{ParticleDataGroup:2024cfk} \\
			\hline	\hline
			  \multirow{8}{*}{\rotatebox{90}{{$J^P=\frac{1}{2}^+$}}} & \multicolumn{4}{l|}{Octet (C = b = 0)}\\ \cline{2-5}
			&$p~(\{ud\}u)$           & $938.175$\footnotemark[2]$^{,}$\footnotemark[3] & $936.94$\footnotemark[2]$^{,}$\footnotemark[3]  & $938.27$    \\	
			&$n~(\{ud\}d)$           & $939.468$\footnotemark[2]$^{,}$\footnotemark[3] & $938.24$\footnotemark[2]$^{,}$\footnotemark[3] & $939.56$  \\
			&$\Lambda^{0}~([us]d)$ & $1112.164$\footnotemark[2]$^{,}$\footnotemark[3] & $1115.68$\footnotemark[2]& $1115.683(6)$  \\	
			&$\Sigma^{+}~(\{us\}u)$  & $1186.796$\footnotemark[2]$^{,}$\footnotemark[3] & $1168.04$ & $1189.37(7)$  \\	
			&$\Sigma^{0}~(\{us\}d)$  & $1189.217$\footnotemark[2]$^{,}$\footnotemark[3] & $1172.25$ & $1192.642(24)$ \\		
			&$\Sigma^{-}~(\{ds\}d)$  & $1196.929$\footnotemark[2]$^{,}$\footnotemark[3] & $1176.19$ & $1197.449(30)$ \\
			&$\Xi^{0}~(\{ss\}u)$     & $1319.609$\footnotemark[2]$^{,}$\footnotemark[3] & $1314.86$\footnotemark[3] & $1314.86(20)$  \\
			&$\Xi^{-}~(\{ss\}d)$     & $1323.264$\footnotemark[2]$^{,}$\footnotemark[3] & $1321.71$\footnotemark[3] & $1321.71(7)$  \\\hline \hline
			\multirow{10}{*}{\rotatebox{90}{{$J^P=\frac{3}{2}^+$}}} & \multicolumn{4}{l|}{Decuplet (C = b = 0)}\\ \cline{2-5}
			&$\Delta^{++}~(\{uu\}u)$ & $1230.813$\footnotemark[2]$^{,}$\footnotemark[3] & $1232.00$\footnotemark[2]$^{,}$\footnotemark[3] & $1230.55(20)$ \\
			\multirow{9}{*}{}&$\Delta^{+}~(\{ud\}u)$  & $1232.673$\footnotemark[2]$^{,}$\footnotemark[3] & $1233.57$\footnotemark[2]$^{,}$\footnotemark[3] & $1234.9(1.4)$ \\
			&$\Delta^{0}~(\{ud\}d)$  & $1233.671$\footnotemark[2]$^{,}$\footnotemark[3] & $1234.86$\footnotemark[2]$^{,}$\footnotemark[3] & $1231.3(6)$ \\
			&$\Delta^{-}~(\{dd\}d)$  & $1234.430$\footnotemark[2]$^{,}$\footnotemark[3] & $1235.89$ & -  \\
			&$\Sigma^{*+}~(\{us\}u)$ & $1378.858$\footnotemark[2]$^{,}$\footnotemark[3] & $1382.74$ & $1382.83(34)$ \\	
			&$\Sigma^{*0}~(\{us\}d)$ & $1378.727$\footnotemark[2]$^{,}$\footnotemark[3] & $1384.03$ & $1383.7(1.0)$  \\		
			&$\Sigma^{*-}~(\{ds\}d)$ & $1386.439$\footnotemark[2]$^{,}$\footnotemark[3] & $1385.04$ & $1387.2(5)$  \\
			&$\Xi^{*0}~(\{ss\}u)$    & $1538.078$\footnotemark[2]$^{,}$\footnotemark[3] & $1529.55$ & $1531.80(32)$ \\
			&$\Xi^{*-}~(\{ss\}d)$    & $1536.713$\footnotemark[2]$^{,}$\footnotemark[3] & $1530.56$ & $1535.0(6)$  \\
			&$\Omega^{-}~(\{ss\}s)$  & $1672.304$\footnotemark[2]$^{,}$\footnotemark[3] & $1672.45$\footnotemark[3] & $1672.45(29)$  \\\hline
		\end{tabular}
    \footnotetext[1]{The quark compositions and effective diquark configurations adopted in Scenario II, where $\{qq\}$ and $[qq]$ represent symmetric and antisymmetric flavor states, respectively.}
	\footnotetext[2]{Used as input in the calculation of constituent quark masses ($m_{i}$).}
	\footnotetext[3]{Used as input in the calculation of hyperfine interaction terms ($b_{ij}$~and~$b_{\{ij\}k}$).}
    \footnotetext[4]{Values in parentheses represent uncertainties.}
	\end{table}
\bibliographystyle{apsrev4-2} 
\bibliography{Ref.bib}

\newpage  
\begin{table}[ht]
	\centering
	\caption{Constituent quark masses and hyperfine interaction terms in Scenario I (in MeV).}
	\label{t1}
		\begin{tabular}{|c|c||c|c|} 	\hline 
			Experimental			& Constituent quark & Experimental & Hyperfine interaction \\
			inputs \cite{ParticleDataGroup:2024cfk} & masses ($m_{i}$) \cite{Mohan:2022sxm} & inputs \cite{ParticleDataGroup:2024cfk} & terms ($b_{ij}$) \cite{Mohan:2022sxm} \\	\hline \hline
			$N, N^{*}$ &$m_{u} = 360.534$ & $N$,$ N^{*}$   &  $b_{uu} = 200.536$ \\
			&$m_{d} =  363.491$  &  $N$, $\Delta^{+} $                  &   $b_{ud} = 197.752$ \\
			&                    &  $N$, $ N^{*}$                        &   $b_{dd} = 193.884$  \\ \hline
			$\Lambda^{0}$                       &$m_{s} =  539.972$  &  $\Xi^{0}$                         &   $b_{us} = 143.129$ \\
			&                    &  $\Xi^{-}$                         &   $b_{ds} = 139.236$ \\
			&                    &  $\Omega^{-}$                      &   $b_{ss} = 70.045$  \\	\hline	
			$\Omega_{c}^{0}$, $\Omega_{c}^{*0}$ &$m_{c} = 1644.878 $ &  $\Xi_{c}^{\prime+}$, $\Xi_{c}^{*+}$ &   $b_{uc} = 42.067$  \\
			&                    &  $\Xi_{c}^{\prime0}$, $\Xi_{c}^{*0}$ &   $b_{dc} = 42.814$  \\
			&                    &  $\Omega_{c}^{0}$                  &   $b_{sc} = 47.133 $ \\
			&                    &  $\Xi_{cc}^{++}$                   &   $b_{cc} = 53.508 $ \\	\hline 
			$\Lambda_b^0$ & $m_{b} = 5043.889$\footnotemark[1] & $\Sigma_{b}^{+}$, $\Sigma_{b}^{*+}$ &   $b_{ub} = 13.173$\footnotemark[1]  \\
			&                    &  $\Sigma_{b}^{-}$, $\Sigma_{b}^{*-}$ &   $b_{db} = 12.733$  \\
			&                    & $\Xi_{b}^{\prime-}$, $\Xi_{b}^{*-}$        &   $b_{sb} = 14.467$  \\
			&                    & $(\frac{m_s}{m_c})b_{sb}$           &   $b_{cb} = 4.749$  \\
			&                    & $(\frac{m_s}{m_b})b_{sb}$     &   $b_{bb} = 1.549$  \\ \hline
	\end{tabular} 
    \footnotetext[1]{$m_i$ and $b_{ij}$ terms in the $SU(5)$ sector is calculated utilizing the input values of the lower flavor sectors given in $2-4$ rows of Table \ref{t1}, which are quoted from our previous work \cite{Mohan:2022sxm}.}
\end{table} 
\begin{table}[ht]
	\centering
	\captionof{table}{Constituent quark masses and hyperfine interaction terms in Scenario II (in MeV).}
	\label{t2}
	\setlength{\tabcolsep}{2pt}
	\begin{tabular}{|c|c||c|c||c|c|} 	\hline 
		Inputs & Constituent & Inputs & Quark-quark & Inputs & Quark-diquark \\
		\cite{ParticleDataGroup:2024cfk} & quark masses ($m_{i}$) & \cite{ParticleDataGroup:2024cfk} & interaction terms ($b_{ij}$) & \cite{ParticleDataGroup:2024cfk} & interaction terms ($b_{\{ij\}k}$)\\	\hline \hline
		$N, N^{*}$\footnotemark[1] &$m_{u} = 352.482$ & $N$,$ N^{*}$   &  $b_{uu} = 104.476$ & $N, N^{*}$ & $b_{\{qq\}u} = 147.249$\\
		&$m_{d} =  353.627$  &  $N, N^{*}$                  &   $b_{ud} = 107.336$ & $N, N^{*}$ & $b_{\{qq\}d} = 147.101$ \\
		&                    &  $\Delta^{-}$                        &   $b_{dd} = 105.788$ & & \\\hline
		
		$\Lambda^{0}, \Sigma^{(*)}$            & $m_{s} =  534.911$  & $\Lambda^{0}, \Sigma^{(*)}$      &   $b_{us} = 171.808$ & $\Lambda^{0}, \Sigma^{(*)}$ & $b_{\{qs\}u} = 96.031 $ \\
		$\Xi^{(*)}, \Omega^{-}$\footnotemark[2]	&                    &  $\Xi^{(*)}, \Omega^{-}$  &   $b_{ds} = 198.077$ & $\Xi^{(*)}, \Omega^{-}$ & $b_{\{qs\}d} = 94.755$ \\
		&                    &    &   $b_{ss} = 26.160$  &  & $b_{\{ss\}u} = 109.234$ \\
		&                    &                                    &                     &  & $b_{\{ss\}d} = 106.724$ \\ 
		&                    &                                    &                     &  & $b_{\{ss\}s} = 61.032$ \\\hline	
		
		$\Xi_{c}^{\prime+}$ & $m_{c} = 1681.305$ & $(\frac{m_{d}m_{s}}{m_{c}^{2}})b_{ds}$  & $b_{cc} = 13.255$ & $\Sigma_{c}^{0}, \Sigma_{c}^{*0}$ & $b_{\{qq\}c} = 32.365$ \\
		&                    &   &  & $\Xi_{c}^{\prime+}, \Xi_{c}^{*+}$ & $b_{\{qs\}c} = 33.450$ \\
		&                    &   &  & $\Omega_{c}^{0}, \Omega_{c}^{*0}$ & $b_{\{ss\}c} = 35.350$ \\
		&                    &   &  & $(\frac{m_{\{qs\}}m_{c}}{m_{\{cc\}}m_{u}})b_{\{qs\}c}$ & $b_{\{cc\}u} = 44.466$ \\
		&                    &                                    &                      & $(\frac{m_{\{qs\}}m_{c}}{m_{\{cc\}}m_{d}})b_{\{qs\}c}$ & $b_{\{cc\}d} = 44.322$ \\
		&                    &                                    &                      & $(\frac{m_{\{qs\}}m_{c}}{m_{\{cc\}}m_{s}})b_{\{qs\}c}$ & $b_{\{cc\}s} = 29.301$ \\
		&                    &                                    &                      & $(\frac{m_{\{qq\}}}{m_{\{cc\}}})b_{\{qq\}c}$ & $b_{\{cc\}c} = 7.048$ \\  \hline 
		
		$\Xi_{b}^{\prime-}$ & $m_{b} = 5007.140$ & $(\frac{m_{c}}{m_{b}})b_{cc}$  & $b_{cb} = 4.451$ & $\Sigma_{b}^{+}, \Sigma_{b}^{*+}$ & $b_{\{qq\}b} = 9.880$ \\
		& & $(\frac{m_{c}}{m_{b}})^2b_{cc}$  & $b_{bb} = 1.494$ & $\Xi_{b}^{\prime-}, \Xi_{b}^{*-}$ & $b_{\{qs\}b} = 10.300$ \\
		& & & & $(\frac{m_{\{qs\}}}{m_{\{ss\}}})b_{\{qs\}b}$ & $b_{\{ss\}b} = 8.977$ \\
		& & & & $(\frac{m_{\{qq\}}m_b}{m_{\{cb\}}m_u})b_{\{qq\}b}$ & $b_{\{cb\}u} = 15.393$ \\
		& & & & $(\frac{m_{\{qq\}}m_b}{m_{\{cb\}}m_d})b_{\{qq\}b}$ & $b_{\{cb\}d} = 15.343$ \\
		& & & & $(\frac{m_{\{qq\}}m_b}{m_{\{cb\}}m_s})b_{\{qq\}b}$ & $b_{\{cb\}s} = 10.144$ \\ 
		& & & & $(\frac{m_{\{qs\}}}{m_{\{cc\}}})b_{\{qs\}b}$ & $b_{\{cc\}b} = 2.871$ \\ 
		& & & & $(\frac{m_{\{qs\}}m_b}{m_{\{bb\}}m_u})b_{\{qs\}b}$ & $b_{\{bb\}u} = 13.705$ \\
		& & & & $(\frac{m_{\{qs\}}m_b}{m_{\{bb\}}m_d})b_{\{qs\}b}$ & $b_{\{bb\}d} = 13.661$ \\
		& & & & $(\frac{m_{\{qs\}}m_b}{m_{\{bb\}}m_s})b_{\{qs\}b}$ & $b_{\{bb\}s} = 9.031$ \\
		& & & & $(\frac{m_{\{qs\}}m_b}{m_{\{bb\}}m_c})b_{\{qs\}b}$ & $b_{\{bb\}c} = 2.873$ \\
		& & & & $(\frac{m_{\{qs\}}}{m_{\{bb\}}})b_{\{qs\}b}$ & $b_{\{bb\}b} = 0.965$ \\ \hline
	\end{tabular} 
	\footnotetext[1]{The values are fixed through minimization using the experimental masses of all the baryons in the $SU(2)$ sector.}
	\footnotetext[2]{The values are fixed through minimization using the experimental masses of all the baryons in the $SU(3)$ sector.}		
\end{table}
\begin{table}[h!]
\centering
\caption{Diquark masses (in MeV) and mass ratios $R = M_A/M_S$ for Scenario I (Sc. I) and Scenario II (Sc. II). Values in parentheses denote ratios for Sc. II. Square brackets $[\,]$ and curly brackets $\{\,\}$ represent scalar ($0^+$) and axial-vector ($1^+$) diquarks, respectively.} \label{t3}
\begin{tabular}{|c|c|c|c|c|c|c|c|c|c|}
\hline
\multicolumn{2}{|c|}{Diquark} & \multirow{2}{*}{Sc. I} & \multirow{2}{*}{Sc. II} & ~PM~ &  ~CIM~ & ~BSE~ & ~$\chi$QM~ & ~RQDM~ & ~RQM~ \\
\multicolumn{2}{|c|}{flavor content} & & & \cite{Ferretti:2019zyh} & \cite{Yin:2019bxe} & \cite{MoosaviNejad:2020nsl} & \cite{Kim:2021ywp} & \cite{Santopinto:2014opa} & \cite{Faustov:2021qqf, Ebert:2011kk} \\
\hline
\multirow{10}{*}{\rotatebox{90}{{Scalar ($J^P=0^+$)}}} & $[ud]$ & $279.08$ & $625.61$ & $691$ & $770$ & - & $725$ & $607$ & $710$ \\
& $[us]$ & $578.47$ & $758.54$ & $886$ & $930$ & - & $942$ & $856$ & $948$ \\
& $[ds]$ & $590.18$ & $739.98$ & - & -  & - & - & - & -\\
& $[uc]$ & $1910.76$ & -    & $2099$ & $2150$ & $2137$ & - & - & -\\
& $[dc]$ & $1912.04$ & -    & - & - & $2195$ & - & - & - \\
& $[sc]$ & $2078.80$ & -    & $2229$ & $2260$ & $2235$ & - & - & - \\
& $[ub]$ & $5374.78$ & -    & $5451$ & $5510$ & $5433$ & - & - & - \\
& $[db]$ & $5378.73$ & -    & - & - & $5498$ & - & - & - \\
& $[sb]$ & $5551.31$ & -    & $5572$ & $5600$ & $5627$ & - & - & - \\
& $[cb]$ & $6678.08$ & $6685.11$ & $6599$ & $6480$ & - & - & - & $6519$ \\
\hline \hline
\multirow{15}{*}{\rotatebox{90}{{Axial-vector($J^P=1^+$)}}} & $\{uu\}$ & $871.47$ & $731.08$ & $840$ & $1060$ & - & $973$ & $963$ & -\\
& $\{ud\}$ & $872.34$ & $732.94$ & - & -  & -  & - & - & $909$ \\
& $\{dd\}$ & $872.40$ & $733.70$ & -  & -  & - & - & - & -\\
& $\{us\}$ & $1007.85$ & $930.34$ & $992$ & $1160$ & - & $1116$ & $1216$ & $1069$ \\
& $\{ds\}$ & $1007.89$ & $938.06$ & - & -  & - & - & - & - \\
& $\{ss\}$ & $1132.48$ & $1076.36$ & $1136$ & $1260$ & - & $1242$ & $1352$ & $1203$ \\
& $\{uc\}$ & $2036.96$ & - & $2138$ & $2240$ & $2175$ & - & - & -\\
& $\{dc\}$ & $2040.48$ & -    & - & - & $2200$ & - & - & -\\
& $\{sc\}$ & $2220.20$ & - & $2264$ & $2340$ & $2270$ & -& - & -\\
& $\{cc\}$ & $3329.89$ & $3365.92$ & $3329$ & $3300$ & - & - & - & $3226$ \\
& $\{ub\}$ & $5414.30$ & - & $5465$ & $5530$ & $5446$ & - & - & -\\
& $\{db\}$ & $5416.93$ & -   & - & - & $5544$ & - & - & -\\
& $\{sb\}$ & $5594.71$ & - & $5585$ & $5620$ & $5638$ & - & - & -\\
& $\{cb\}$ & $6692.33$ & $6689.56$ & $6611$ & $6500$ & - & - & - & $6526$ \\
& $\{bb\}$ & $10088.90$ & $10014.70$ & $9845$ & $9680$ & - & - & - & $9778$ \\
\hline
\multicolumn{2}{|c|}{Ratios $ (R)$}&\multicolumn{8}{c|}{$R_{ud}=3.13(1.17)$, $R_{us}=1.74(1.23)$, $R_{uc}=1.07$, $R_{sc}=1.07$,}\\
\multicolumn{2}{|c|}{across flavors}&\multicolumn{8}{c|}{$R_{ub}=1.01$, $R_{sb}=1.01$, $R_{cb}=1.00(1.00)$}\\\hline
\end{tabular}
\end{table}
\FloatBarrier
\begin{figure}[]
    \centering
    \includegraphics[width=1.0\linewidth]{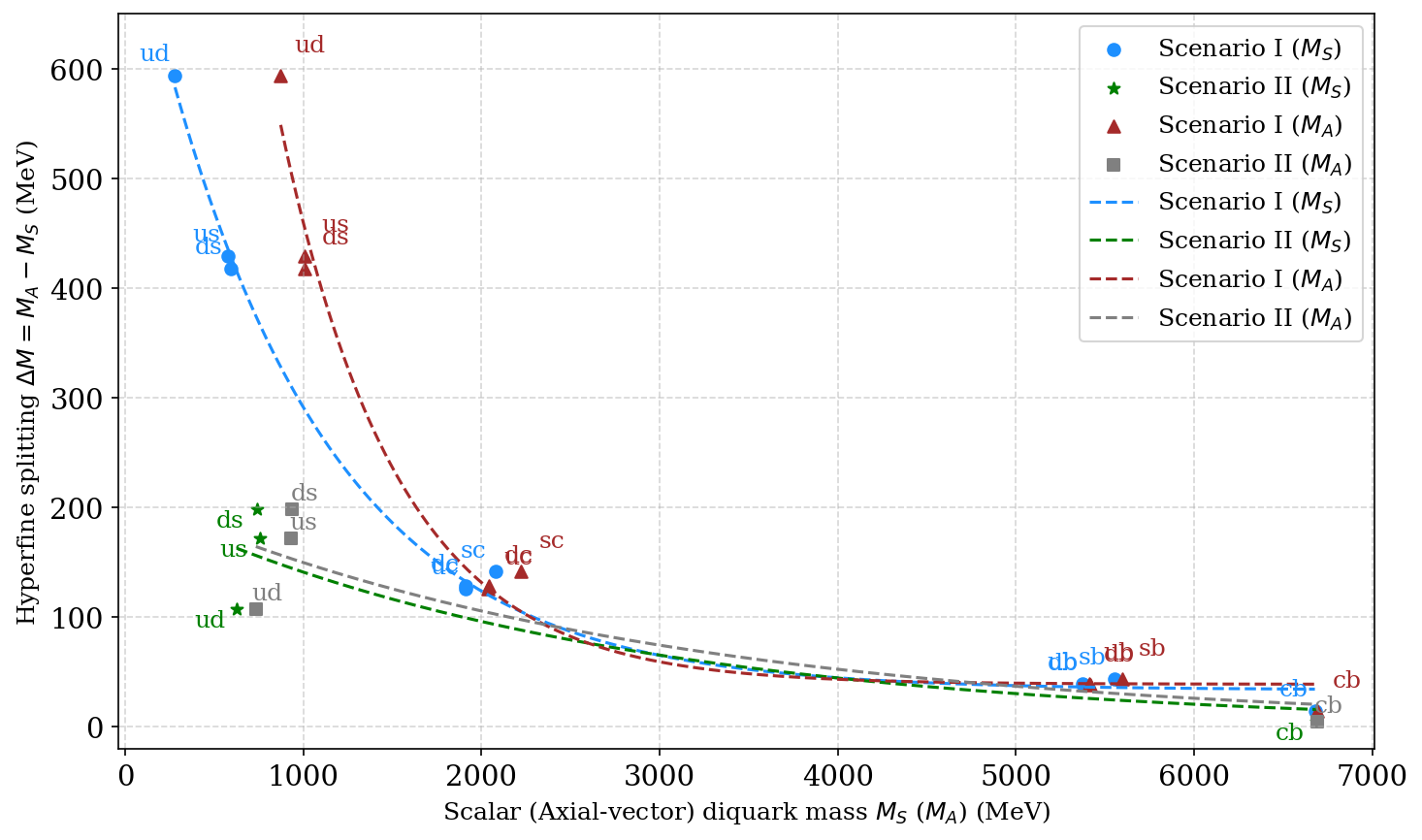}
    \caption{Variation of hyperfine splitting, $\Delta M= M_A - M_S$ in scenarios I and II with corresponding diquark masses.}
    \label{Fig1}
\end{figure}
\begin{table}[ht]
	\centering
	\captionof{table}{Binding energy (in MeV).}
	\label{t4}
	\begin{tabular}{|c|c|c|c|c|}\hline
		Experimental & \multicolumn{2}{c|}{\multirow{2}{*}{$BE(Q\overline{Q}^{\prime})$}} & \multicolumn{2}{c|}{\multirow{2}{*}{$BE(QQ^{\prime})$}}\\ 
        inputs \cite{ParticleDataGroup:2024cfk} & \multicolumn{2}{c|}{} & \multicolumn{2}{c|}{} \\ \hline \hline
		$D_s^{*+}, D_s^{+}$ & $c\overline{s}$ & $-78.55$ & $cs$ & $-39.28$ \\
		$J/\Psi, \eta_c$ & $c\overline{c}$ & $-262.36$ & $cc$ & $-131.18$ \\ \hline
		$B_s^{*0}, B_s^{0}$ & $\overline{b}s$ & $-93.57$ & $bs$ & $-46.79$ \\
		$B_c^{*+}$\footnote{The $B_c^{*+}$ mass of $6331$ MeV from LQCD \cite{Mathur:2018epb} is used to compute the binding energy term $BE(\overline{b}c)$.}, $B_c^{+}$ & $\overline{b}c$ & $-356.26$ & $bc$ & $-178.13$ \\
		$\Upsilon, \eta_b$ & $\overline{b}b$ & $-570.29$ & $bb$ & $-285.14$ \\ \hline
	\end{tabular}
\end{table}	
\begin{figure}[]
    \centering
    \includegraphics[width=0.95\linewidth]{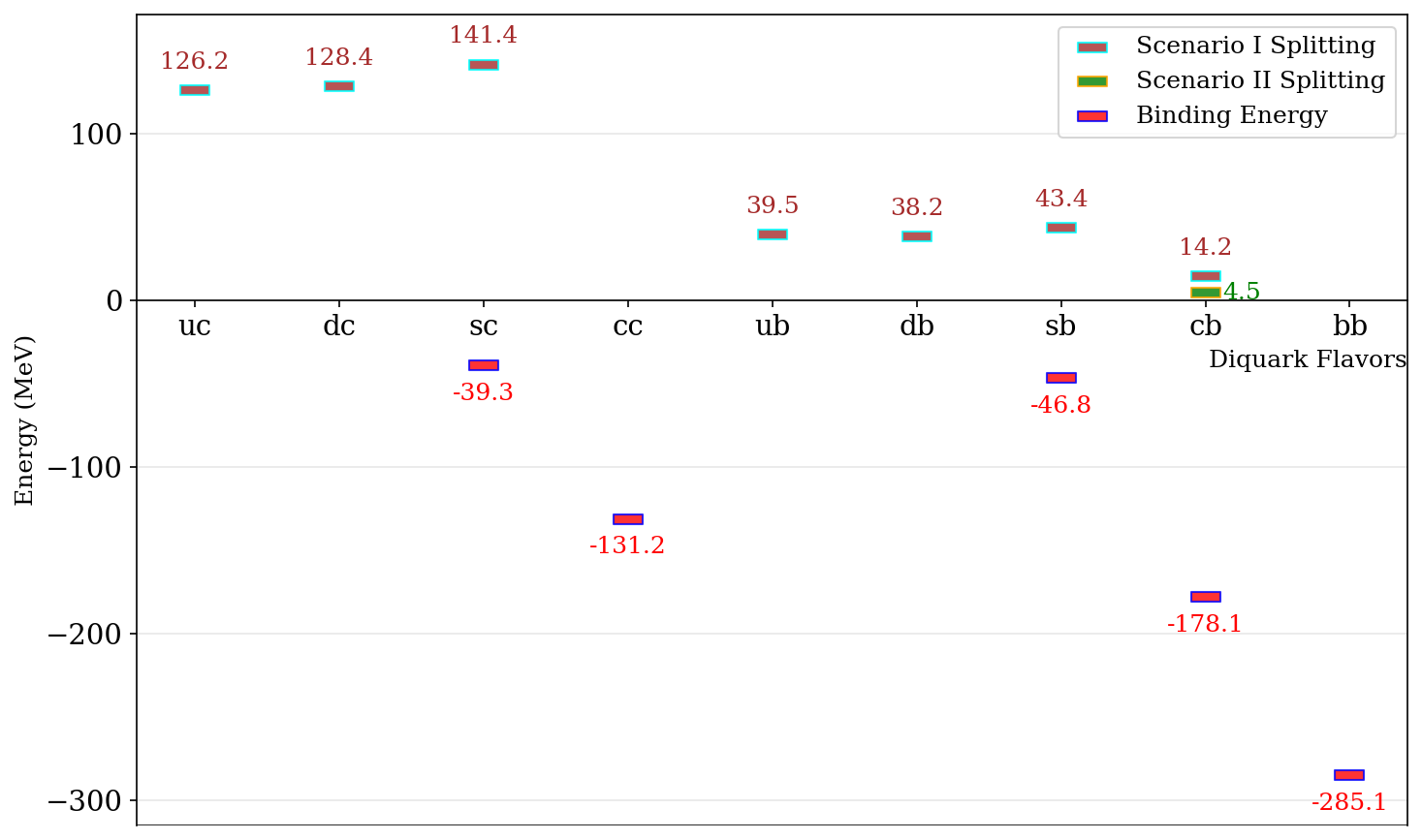}
    \caption{Lower half illustrates the diquark binding energies, while the upper half shows the hyperfine mass splittings for different diquark flavors in scenarios I and II.}
    \label{Fig2}
\end{figure}
\begin{table}[ht]
	\centering
	\captionof{table}{Masses of charm baryons in quark-diquark model (in MeV).} 
	\label{t5}
	\scalebox{0.9}{\begin{tabular}{|c|c|c|c|c|c|c|c|c|c|c|c|}	\hline  
		 \multicolumn{2}{|c|}{} & \multicolumn{4}{c|}{This work} & \multirow{2}{*}{CIM} & \multirow{2}{*}{BSE\footnotemark[4]} & \multirow{2}{*}{HB$\chi$PT} & \multirow{2}{*}{HCQM} & \multirow{2}{*}{BM} & \multirow{2}{*}{PDG} \\ \cline{3-6}
	 \multicolumn{2}{|c|}{Baryons\footnotemark[1]} &\multicolumn{2}{c|}{with $BE$} & \multicolumn{2}{c|}{without $BE$} & \multirow{2}{*}{\cite{Yin:2019bxe}} & \multirow{2}{*}{\cite{Farhadi:2023ucs}} &\multirow{2}{*}{\cite{Jiang:2014ena}} &\multirow{2}{*}{\cite{Shah:2016mig, Shah:2016vmd, Shah:2017liu, Shah:2017jkr}} & \multirow{2}{*}{\cite{Zhang:2021yul}} & \multirow{2}{*}{\cite{ParticleDataGroup:2024cfk}} \\
		\cline{3-6}
		\multicolumn{2}{|c|}{} & Sc. II & Sc. I & Sc. II & Sc. I \cite{Mohan:2022sxm} & & & & & & \\
		\hline	\hline
		\multirow{15}{*}{\rotatebox{90}{{$J^P=\frac{1}{2}^+$}}} 
        & \multicolumn{11}{l|}{Antitriplet (C = 1)} \\\cline{2-12}
		& $\Lambda_c^{+}~([ud]c)$      & $2306.91$ & $2220.59$ & $2306.91$ & $2220.59$ & $2500$ & $2622.0$ & $2286.46$\footnotemark[5] & $2286$\footnotemark[5] & $2270$ &  $2286.46(14)$ \\
		& $\Xi_c^{+}~([us]c)$          & $2439.84$ & $2398.76$ & $2439.84$ & $2438.04$ & $2660$ & $2423.2$ & $2467.80$\footnotemark[5] & $2467$\footnotemark[5] & $2436$ & $2467.71(23)$ \\	
		& $\Xi_c^{0}~([ds]c)$          & $2421.29$ & $2404.63$ & $2421.29$ & $2443.91$ & -  & $2462.8$ & $2470.88$\footnotemark[5] & $2470$\footnotemark[5] & - & $2470.44(28)$ \\\cline{2-12}

        & \multicolumn{11}{l|}{Sextet (C = 1)} \\\cline{2-12}
		& $\Sigma_c^{++}~(\{uu\}c)$      & $2380.02$ & $2374.01$ & $2380.02$ & $2374.01$ & $2530$ & $2445.7$ & $2454.02$\footnotemark[5] & $2454$\footnotemark[5] & $2411$ & $2453.97(14)$ \\
		& $\Sigma_c^{+}~(\{ud\}c)$       & $2381.88$ & $2375.90$ & $2381.88$ & $2375.90$ & -      & $2415.5$ & $2452.90$\footnotemark[5] & $2452$\footnotemark[5] & - &  $2452.65^{+0.22}_{-0.16}$ \\
		& $\Sigma_c^{0}~(\{dd\}c)$       & $2382.64$ & $2377.52$ & $2382.64$\footnotemark[2] & $2377.52$ & - & $2486.8$ &  $2453.76$\footnotemark[5] & $2453$\footnotemark[5] & - & $2453.75(14)$ \\	
		& $\Xi_c^{\prime+}~(\{us\}c)$    & $2578.20$ & $2497.29$ & $2578.20$\footnotemark[2]$^{,}$\footnotemark[3] & $2536.57$\footnotemark[2] & $2680$ & - & $2572.66$ & -      & $2544$ & $2578.2(5)$ \\
        & $\Xi_c^{\prime0}~(\{ds\}c)$    & $2585.91$ & $2498.90$ & $2585.91$ & $2538.18$\footnotemark[2] & - & - & $2570.40$ & -      & - & $2578.7(5)$ \\
		& $\Omega_c^{0}~(\{ss\}c)$       & $2722.32$ & $2616.64$ & $2722.32$\footnotemark[2] & $2695.20$\footnotemark[2]$^{,}$\footnotemark[3] & $2830$ & $2745.6$ & $2695.20$\footnotemark[5] & $2695$\footnotemark[5] & $2680$ & $2695.3(4)$ \\ \cline{2-12}
		
		& \multicolumn{11}{l|}{Triplet (C = 2)} \\ \cline{2-12}
		& $\Xi_{cc}^{++}~(\{cc\}u)$      & $3542.76$ & $3490.42$ & $3673.94$ & $3621.60$\footnotemark[2] & $3720$ & $3598$ & $3591$ \cite{Yao:2018ifh} & $3511$ & $3604$ & $3621.6(4)$ \\
		& $\Xi_{cc}^{+}~(\{cc\}d)$       & $3544.05$ & $3492.63$ & $3675.23$ & $3623.81$ & - & $3493$ & - & $ 3520$ & - & $3623.0(1.4)$ \cite{LHCb:2021eaf}\footnotemark[6]\\
		& $\Omega_{cc}^{+}~(\{cc\}s)$    & $3740.35$ & $3586.23$ & $3871.53$ & $3795.97$ & $3900$ & $3601$ & $3657$ \cite{Yao:2018ifh} & $3650$ & $3726$ & $3733(13)$\footnotemark[7] \\ \hline \hline
		
		\multirow{13}{*}{\rotatebox{90}{{$J^P=\frac{3}{2}^+$}}}
        & \multicolumn{11}{l|}{Sextet (C = 1)} \\ \cline{2-12}
		& $\Sigma_c^{*++}~(\{uu\}c)$     & $2444.75$ & $2437.11$ & $2444.75$ & $2437.11$ & $2560$ & $2483.1$ & $2518.40$\footnotemark[5] & $2530$ & $2512$ & $2518.41(22)$ \\
		& $\Sigma_c^{*+}~(\{ud\}c)$      & $2446.61$ & $2439.56$ & $2446.61$ & $2439.56$ & - & $2434.5$ & $2517.50$\footnotemark[5] & $2501$ & - & $2517.4^{+0.7}_{-0.5}$ \\
		& $\Sigma_c^{*0}~(\{dd\}c)$      & $2447.37$ & $2441.74$ & $2447.37$\footnotemark[2] & $2441.74$ & - & $2495.9$ & $2518.0$\footnotemark[5]  & $2529$ & - & $2518.48(21)$ \\	
		& $\Xi_c^{*+}~(\{us\}c)$         & $2645.10$ & $2564.19$ & $2645.10$\footnotemark[2] & $2603.47$\footnotemark[2] & $2700$ & $2554.1$ & $2636.83$ & $2619$ & $2636$ & $2645.10(30)$ \\	
		& $\Xi_c^{*0}~(\{ds\}c)$         & $2652.81$ & $2566.36$ & $2652.81$ & $2605.64$\footnotemark[2] & - & $2466.3$ & $2633.71$ & $2610$ & - & $2645.7^{+0.6}_{-0.7}$ \\		
		& $\Omega_c^{*0}~(\{ss\}c)$      & $2793.02$ & $2687.34$ & $2793.02$\footnotemark[2] & $2765.90$\footnotemark[3] & $2850$ & $2796.3$ & $2765.90$\footnotemark[5] & $2745$ & $2764$ & $2766.0^{+0.9}_{-1.0}$ \\ \cline{2-12}		
		
        & \multicolumn{11}{l|}{Triplet (C = 2)} \\ \cline{2-12}
		& $\Xi_{cc}^{*++}~(\{cc\}u)$     & $3631.69$ & $3553.52$ & $3762.87$ & $3684.70$ & $3750$ & $3671$ & - & $3687$ & $3714$ & $3703(33)$\footnotemark[7] \\
		& $\Xi_{cc}^{*+}~(\{cc\}d)$      & $3632.69$ & $3556.85$ & $3763.87$ & $3688.03$ & - & $3621$ &  - & $3695$ & - & - \\
		& $\Omega_{cc}^{*+}~(\{cc\}s)$   & $3798.96$ & $3656.93$ & $3930.14$ & $3866.67$ & $3940$ & $3712$ & - & $3810$ & $3820$ & $3793(30)$\footnotemark[7] \\ \cline{2-12}
		
		& \multicolumn{11}{l|}{Singlet (C = 3)} \\ \cline{2-12}
		& $\Omega_{ccc}^{*++}~(\{cc\}c)$ & $4923.10$ & $4581.22$ & $5054.28$ & $4974.76$ & $5000$ & - & - & $4806$ & - & $4817(12)$\footnotemark[7] \\ \hline
	\end{tabular}  }
	\footnotetext[1]{The quark compositions and effective diquark configurations adopted in Scenario II, where $\{qq\}$ and $[qq]$ represent symmetric and antisymmetric flavor states, respectively.}
	\footnotetext[2]{Used as input in the calculation of hyperfine interaction terms.}
    \footnotetext[3]{Used as input in the calculation of constituent quark masses ($m_{i}$).}
    \footnotetext[4]{Hadron mass differences and non-strange baryon magnetic moments are used as inputs for the fit.}
    \footnotetext[5]{Used as input in respective models.}
    \footnotetext[6]{PDG average is 3518.9(9) MeV.}
	\footnotetext[7]{Results are taken from LQCD \cite{Bahtiyar:2020uuj}.}
\end{table}
\begin{table}[ht]
	\centering
	\captionof{table}{Masses of singly bottom baryons in quark-diquark model (in MeV).} 
	\label{t6}
	\scalebox{0.90}{\begin{tabular}{|c|c|c|c|c|c|c|c|c|c|c|c|c|}	\hline  
		\multicolumn{2}{|c|}{} &\multicolumn{4}{c|}{This work} & \multirow{2}{*}{BSE\footnotemark[4]} & \multirow{2}{*}{$\chi$QM} & \multirow{2}{*}{RQM} & \multirow{2}{*}{BM} & \multirow{2}{*}{QCDSR} & \multirow{2}{*}{NRQM} & \multirow{2}{*}{PDG} \\ \cline{3-6}
		\multicolumn{2}{|c|}{Baryons\footnotemark[1]} &\multicolumn{2}{c|}{with $BE$} & \multicolumn{2}{c|}{without $BE$} &\multirow{2}{*}{\cite{Farhadi:2023ucs}} & \multirow{2}{*}{\cite{Kim:2021ywp}} & \multirow{2}{*}{\cite{Ebert:2011kk}} & \multirow{2}{*}{\cite{Zhang:2021yul}} & \multirow{2}{*}{\cite{Liu:2007fg}}& \multirow{2}{*}{\cite{Ortiz-Pacheco:2023kjn}} & \multirow{2}{*}{\cite{ParticleDataGroup:2024cfk}}\\
		\cline{3-6}
		\multicolumn{2}{|c|}{} & Sc. II & Sc. I & Sc. II & Sc. I & & & & & & & \\ \hline	\hline
		\multirow{11}{*}{\rotatebox{90}{{$J^P=\frac{1}{2}^+$}}} &
        \multicolumn{12}{l|}{Antitriplet (b = 1)}\\ \cline{2-13}
		& $\Lambda_b^{0}~([ud]b)$      & $5632.75$ & $5619.60$ & $5632.75$ & $5619.60$\footnotemark[3] & $5510.8$ & $5620$ & $5620$ & $5648$ & $5637^{+68}_{-56}$ & $5615$ & $5619.57(16)$ \\
		& $\Xi_b^{0}~([us]b)$          & $5765.68$ & $5790.26$ & $5765.68$ & $5837.05$ & $5719.7$ & $5796$ & $5803$ & $5805$ & $5780^{+73}_{-68}$ & $5812$ & $5791.7(4)$ \\	
		& $\Xi_b^{-}~([ds]b)$          & $5747.12$ & $5796.14$ & $5747.12$ & $5842.93$ & $5872.0$ & - & - & - & - & - & $5797.0(4)$ \\\cline{2-13} 
		
		& \multicolumn{12}{l|}{Sextet (b = 1)}\\ \cline{2-13}
		& $\Sigma_b^{+}~(\{uu\}b)$      & $5728.34$ & $5801.92$ & $5728.34$\footnotemark[2] & $5801.92$\footnotemark[2] & $5842.0$ & $5810$ & $5808$ & $5835$ & $5809^{+82}_{-76}$ & $5810$ & $5810.56(25)$ \\
		& $\Sigma_b^{0}~(\{ud\}b)$       & $5730.20$ & $5804.40$ & $5730.20$ & $5804.40$ & $5815.4$ & - & - & - & - & - & $5813.10(18)$\footnotemark[5] \\
		& $\Sigma_b^{-}~(\{dd\}b)$       & $5730.96$ & $5806.61$ & $5730.96$ & $5806.61$\footnotemark[2] & $5802.0$ & - & - & - & - & - & $5815.64(27)$  \\	
		& $\Xi_b^{\prime0}~(\{us\}b)$    & $5927.19$ & $5919.57$ & $5927.19$ & $5966.36$ & - & - & - & - & - & - & - \\
		& $\Xi_b^{\prime-}~(\{ds\}b)$    & $5934.90$ & $5921.77$ & $5934.90$\footnotemark[2]$^{,}$\footnotemark[3] & $5968.56$\footnotemark[2] & - & $5934$ & $5936$ & $5956$ & $5903^{+81}_{-79}$ & $5935$ & $5934.9(4)$  \\
		& $\Omega_b^{-}~(\{ss\}b)$       & $6074.53$ & $6033.30$ & $6074.53$ & $6126.88$ & $5868.4$ & $6047$ & $6064$ & $6080$ & $6036\pm81$ & $6078$ & $6045.8(8)$ \\ \hline \hline
        \multirow{7}{*}{\rotatebox{90}{{$J^P=\frac{3}{2}^+$}}} &
        \multicolumn{12}{l|}{Sextet (b = 1)}\\ \cline{2-13}
		& $\Sigma_b^{*+}~(\{uu\}b)$     & $5748.10$ & $5821.68$ & $5748.10$\footnotemark[2] & $5821.68$\footnotemark[2] & $5868.2$ & $5829$ & $5834$ & $5872$ & $5835^{+82}_{-77}$ & $5832$ & $5830.32(27)$ \\
		& $\Sigma_b^{*0}~(\{ud\}b)$      & $5749.96$ & $5823.83$ & $5749.96$ & $5823.83$ & $5842.5$ & - & - & - & - & - & $5832.53(20)$\footnotemark[5] \\
		& $\Sigma_b^{*-}~(\{dd\}b)$      & $5750.72$ & $5825.71$ & $5750.72$ & $5825.71$\footnotemark[2] & $5821.0$ & - & - & - & - & -& $5834.74(30)$ \\	
		& $\Xi_b^{*0}~(\{us\}b)$         & $5947.79$ & $5940.30$ & $5947.79$ & $5987.09$ & $5897.8$ & $5952$ & - & $5991$ & $5929^{+83}_{-79}$ & $5957$ & $5952.3(6)$ \\	
		& $\Xi_b^{*-}~(\{ds\}b)$         & $5955.50$ & $5942.17$ & $5955.50$\footnotemark[2] & $5988.96$\footnotemark[2] & $5924.0$ & - & $5963$ & - & - & - & $5955.5(4)$ \\	
		& $\Omega_b^{*-}~(\{ss\}b)$      & $6092.48$ & $6055.00$ & $6092.48$ & $6148.58$ & $6075.0$ & $6064$ & $6088$ & $6112$ & $6063^{+83}_{-82}$ & $6100$ & $6019(20)$\footnotemark[6] \\ \hline		
	\end{tabular} }
	\footnotetext[1]{The quark compositions and effective diquark configurations adopted in Scenario II, where $\{qq\}$ and $[qq]$ represent symmetric and antisymmetric flavor states, respectively.}
	\footnotetext[2]{Used as input in the calculation of hyperfine interaction terms.}
    \footnotetext[3]{Used as input in the calculation of constituent quark masses ($m_{i}$).}
    \footnotetext[4]{Hadron mass differences and non-strange baryon magnetic moments are used as inputs for the fit.}
    \footnotetext[5]{Calculated according to the definition $M_{\Sigma_b^{(*)0}} = \frac{1}{2}\big(M_{\Sigma_b^{(*)+}}^{Expt} + M_{\Sigma_b^{(*)-}}^{Expt}\big)$.}
    \footnotetext[6]{Results are taken from LQCD \cite{Mohanta:2019mxo}.}
\end{table}
\begin{table}[ht]
	\centering
	\captionof{table}{Masses of charmed bottom, doubly bottom, and triply heavy baryons in quark-diquark model (in MeV).} 
	\label{t7}
	\scalebox{0.9}{\begin{tabular}{|c|c|c|c|c|c|c|c|c|c|c|c|c|}	\hline  
		\multicolumn{2}{|c|}{} & \multicolumn{4}{c|}{This work} & \multirow{2}{*}{BSE\footnotemark[2]} & \multirow{2}{*}{RQM} & \multirow{2}{*}{HCQM} & \multirow{2}{*}{BM} & \multirow{2}{*}{QCDSR} & \multirow{2}{*}{NRQM} & \multirow{2}{*}{LQCD} \\ \cline{3-6}
		 \multicolumn{2}{|c|}{Baryons\footnotemark[1]} & \multicolumn{2}{c|}{with $BE$} & \multicolumn{2}{c|}{without $BE$} &\multirow{2}{*}{\cite{Farhadi:2023ucs}} & \multirow{2}{*}{\cite{Faustov:2021qqf, Ebert:2002ig}} & \multirow{2}{*}{\cite{Shah:2016vmd, Shah:2017jkr, Shah:2017liu}} & \multirow{2}{*}{\cite{Zhang:2021yul}} & \multirow{2}{*}{\cite{ShekariTousi:2024mso, Aliev:2014lxa, Aliev:2012iv, Aliev:2012tt}} & \multirow{2}{*}{\cite{Roberts:2007ni}} & \multirow{2}{*}{\cite{Mohanta:2019mxo}} \\
		\cline{3-6}
		\multicolumn{2}{|c|}{} & Sc. II & Sc. I & Sc. II & Sc. I & & & & & & & \\
		\hline	\hline
		\multirow{11}{*}{\rotatebox{90}{{$J^P=\frac{1}{2}^+$}}} & \multicolumn{12}{l|}{Antitriplet (C = 1, b = 1)}\\ \cline{2-13} 
		& $\Xi_{cb}^{+}~([cb]u)$      & $6859.46$ & $6839.62$ & $7037.59$ & $7017.75$  & $6976$ & $6963$ & $6914$ & $6953$ & $6730^{+140}_{-130}$ & $7011$ & $ 6787(12)$ \\
		& $\Xi_{cb}^{0}~([cb]d)$       & $6860.60$ & $6842.02$ & $7038.73$ & $7020.15$ & $6915$ & - & $6920$ & - & - & - & - \\
		& $\Omega_{cb}^{0}~([cb]s)$    & $7041.89$ & $6929.19$ & $7220.02$ & $7193.39$ & $6916$ & $7116$ & $7136$ & $7064$ & $6770^{+130}_{-120}$ & $7136$ & $6893(16)$ \\ \cline{2-13} 
		
		& \multicolumn{12}{l|}{Triplet (C = 1, b = 1)}\\ \cline{2-13}
		& $\Xi_{cb}^{\prime+}~(\{cb\}u)$      & $6848.52$ & $6872.73$ & $7026.65$ & $7050.86$ & $6680$ & $6933$ & - & $7015$ & $6810\pm110$ & $7047$ & $ 6843(19)$ \\ 
        & $\Xi_{cb}^{\prime0}~(\{cb\}d)$       & $6849.71$ & $6876.09$ & $7027.84$ & $7054.22$ & $6720$ & - & - & - & - & - & - \\
		& $\Omega_{cb}^{\prime0}~(\{cb\}s)$    & $7036.19$ & $6966.71$ & $7214.32$ & $7230.91$ & $6820$ & $7088$ & - & $7116$ & $6820\pm120$ & $7165$ & $6946(17)$ \\ \cline{2-13}
		& \multicolumn{12}{l|}{Singlet (C = 2, b = 1)} \\ \cline{2-13} 
        & $\Omega_{ccb}^{+}~(\{cc\}b)$    & $8239.01$ & $7854.83$ & $8370.19$ & $8342.27$ & $7460$ & $7984$ & - & - & $8500\pm120$ & $8245$ & $ 7797(11)$ \\ \cline{2-13}
		& \multicolumn{12}{l|}{Triplet (b = 2)} \\ \cline{2-13}
		& $\Xi_{bb}^{0}~(\{bb\}u)$      & $10068.30$ & $10150.40$ & $10353.40$ & $10435.50$ & $10106$ & $10202$ & $10312$ & $10311$ & $9970\pm190$ & $10340$ & $10091(17)$ \\
		& $\Xi_{bb}^{-}~(\{bb\}d)$       & $10069.50$ & $10153.80$ & $10354.60$ & $10438.90$ & $10093$ & - & $10317$ & - & - & - & - \\
		& $\Omega_{bb}^{-}~(\{bb\}s)$    & $10255.40$ & $10235.00$ & $10540.50$ & $10613.70$ & $10193$ & $10359$ & $10446$ & $10408$ & $9980\pm180$ & $10454$ & $10190(17)$ \\ \cline{2-13}
		& \multicolumn{12}{l|}{Singlet (C = 1, b = 2)} \\ \cline{2-13}
        & $\Omega_{cbb}^{0}~(\{bb\}c)$    & $11408.00$ & $11086.90$ & $11693.10$ & $11728.30$ & $12110$ & $11198$ & - & - & $11730\pm160$ & $11535$ & $11060(23)$ \\ \hline \hline
		\multirow{7}{*}{\rotatebox{90}{{$J^P=\frac{3}{2}^+$}}} & \multicolumn{12}{l|}{Triplet (C = 1, b = 1)} \\ \cline{2-13}
		& $\Xi_{cb}^{*+}~(\{cb\}u)$     & $6879.30$ & $6886.17$ & $7057.43$ & $7064.30$ & $9973$ & $6980$ & $6980$ & $7044$ & $7250\pm200$ & $7074$ & $ 6835(20)$ \\
		& $\Xi_{cb}^{*0}~(\{cb\}d)$      & $6880.40$ & $6889.20$ & $7058.53$ & $7067.33$ & $6943$ & - & $6986$ & - & - & - & - \\
		& $\Omega_{cb}^{*0}~(\{cb\}s)$   & $7056.48$ & $6981.13$ & $7234.61$ & $7245.33$ & $6997$ & $7130$ & $7187$ & $7142$ & $7300\pm200$ & $7187$ & $6930(19)$ \\ \cline{2-13}
		& \multicolumn{12}{l|}{Singlet (C = 2, b = 1)} \\ \cline{2-13}
        & $\Omega_{ccb}^{*+}~(\{cc\}b)$    & $8244.75$ & $7861.96$ & $8375.93$ & $8349.40$ & - & $7999$ & - & - & $8070\pm100$ & $8265$ & $7807(11)$ \\ \cline{2-13}
		& \multicolumn{12}{l|}{Triplet (b = 2)} \\ \cline{2-13}
		& $\Xi_{bb}^{*0}~(\{bb\}u)$     & $10095.70$ & $10170.16$ & $10380.80$ & $10455.30$ & $10141$ & $10237$ & $10335$ & $10360$ & $10400\pm1000$ & $10367$ & $10103(24)$ \\
		& $\Xi_{bb}^{*-}~(\{bb\}d)$      & $10096.80$ & $10172.86$ & $10381.90$ & $10458.00$ & $10242$ & - & $10340$ & - & - & - & - \\
		& $\Omega_{bb}^{*-}~(\{bb\}s)$   & $10273.50$ & $10256.68$ & $10558.60$ & $10635.40$ & $10220$ & $10389$ & $10467$ & $10451$ & $10500\pm200$ & $10486$ & $10203(22)$ \\ \cline{2-13}
		& \multicolumn{12}{l|}{Singlet (C = 1, b = 2)} \\ \cline{2-13}
        & $\Omega_{cbb}^{*0}~(\{bb\}c)$    & $11413.70$ & $11094.00$ & $11698.80$ & $11735.40$ & - & $11217$ & - & - & $11350\pm150$ & $11554$ & $11081(21)$ \\ \cline{2-13}
		& \multicolumn{12}{l|}{Singlet (b = 3)} \\ \cline{2-13}
		& $\Omega_{bbb}^{*-}~(\{bb\}b)$ & $14737.70$ & $14277.38$ & $15022.80$ & $15132.80$ & - & $14468$ & $14496$ & - & $14300\pm200$ & $14834$ & $14403(7)$ \\ \hline
	\end{tabular} }
	\footnotetext[1]{The quark compositions and effective diquark configurations adopted in Scenario II, where $\{qq\}$ and $[qq]$ represent symmetric and antisymmetric flavor states, respectively.}
	\footnotetext[2]{Hadron mass differences and non-strange baryon magnetic moments are used as inputs for the fit.}
\end{table}
\end{document}